# Analysis of high-order velocity moments in a strained channel flow.

Svetlana V. Poroseva[1] and Scott M. Murman[2]

[1]Mechanical Engineering Department, University of New Mexico, Albuquerque, New Mexico, U.S.A.
[2]NASA Ames Research Center, Moffett Field, California, U.S.A.

## Abstract

In the current study, model expressions for fifth-order velocity moments obtained from the truncated Gram-Charlier series expansions model for a turbulent flow field probability density function are validated using data from direct numerical simulation (DNS) of a planar turbulent flow in a strained channel. Simplicity of the model expressions, the lack of unknown coefficients, and their applicability to non-Gaussian turbulent flows make this approach attractive to use for closing turbulent models based on the Reynolds-averaged Navier-Stokes equations. The study confirms validity of the model expressions. It also shows that the imposed flow deformation improves an agreement between the model and DNS profiles for the fifth-order moments in the flow buffer zone including when the flow reverses its direction. The study reveals sensitivity of particularly odd velocity moments to the grid resolution. A new length scale is proposed as a criterion for the grid generation near the wall and in the other flow areas dominated by high mean velocity gradients when higher-order statistics have to be collected from DNS.

## 1. Introduction

The idea of using the Gram-Charlier series expansions (GCSE) based on Hermite polynomials for modeling a joint probability density function (PDF) of a turbulent flow field belongs to Kampé de Fériet [1]. In [1], the joint PDF, $P(s_1, s_2)$, for two random variables, $s_1$ and $s_2$, is defined as

$$P(s_1,s_2) = P_0(s_1,s_2)\sum_{l+k=0}^{\infty} A_{l,k} H_{l,k}(s_1,s_2), \text{ where } A_{l,k} = \frac{1}{l!k!}\langle G_{l,k}(s_1,s_2)\rangle, \quad (1)$$

where $P_0(s_1, s_2)$ is the Gaussian distribution, and $H_{l,k}(s_1, s_2)$ and $G_{l,k}(s_1, s_2)$ are two adjunct Hermite polynomials [2] (Appendix A). Hereafter, the angular brackets $\langle \ldots \rangle$ symbolize averaging over the statistical ensemble of all possible realizations. In a turbulent flow, components of the turbulent velocity are used to define $s_1$ and $s_2$: $s_1 = u_1/\sigma_1$ and $s_2 = u_2/\sigma_2$, where $u_1$ and $u_2$ can be any component of the turbulent velocity vector $\vec{u} = (u, v, w)$ (in the Cartesian coordinates), and $\sigma_\alpha = \sqrt{\langle u_\alpha^2 \rangle}, \alpha = 1, 2$ (no summation over repeated indices) is their standard deviations [2]. When $A_{0,0} = 1$ and all other coefficients $A_{l,k} = 0$, expression (1) is reduced to the Gaussian distribution $P_0(s_1, s_2)$.

Hermite polynomials can explicitly be expressed in terms of random variables $s_1$ and $s_2$ [1-3]. In Appendix A, these expressions are provided for $G_{l,k}(s_1, s_2)$ at $l + k \leq 4$ (A6)-(A7). When using such expressions in the definition of $A_{l,k}$ along with the definitions of $s_1$ and $s_2$ in a turbulent flow, one obtains expressions for $A_{l,k}$ in terms of the standardized statistical velocity moments $R^{m,n} = \langle u_1^m u_2^n \rangle / \left(\sigma_1^m \sigma_2^n\right)$. In Appendix A, these expressions are provided for $A_{l,k}$ at $l + k \leq 4$ (Eq. (A8)).

Truncation of (1) to various orders leads to the joint PDF models of different accuracy. The procedure is equivalent to imposing the condition $A_{l,k} = 0$ on the coefficients with $l + k > N$, where *N* is the truncation order. For example, if $N = 3$, then, all $A_{l,k}$ with $l + k \geq 4$, are equal to zero. Application of this condition to expressions (A8) with $l + k = 4$ yields:

$$A_{4,0} = \frac{R^{4,0} - 3}{24} = 0, \qquad A_{3,1} = \frac{R^{3,1} - 3R}{6} = 0, \qquad A_{2,2} = \frac{R^{2,2} - 2R^2 - 1}{4} = 0, \quad (2)$$

$$A_{1,3} = \frac{R^{1,3} - 3R}{6} = 0, \qquad A_{0,4} = \frac{R^{0.4} - 3}{24} = 0,$$

which leads to conditions on the standardized velocity moments:

$$R^{4,0} - 3 = 0, R^{3,1} - 3R = 0, R^{2,2} - 2R^2 - 1 = 0, R^{1,3} - 3R = 0, R^{0.4} - 3 = 0. \qquad (3)$$

In (2) and (3), $R = R^{1,1}$. Similar expressions can be obtained for higher-order standardized moments by applying the same procedure to $A_{l,k}$ with $l + k > 4$ [4,5]. Validation of a PDF model based on GCSE truncated at a given *N* includes validation of the condition $A_{l,k} = 0$ at $l + k > N$ or equivalently, the corresponding conditions on the standardized velocity moments.

At *N* =2, for example, the condition $A_{l,k} = 0$ at $l + k > 2$ is satisfied when the odd velocity moments are equal to zero and all even velocity moments are functions of the second-order velocity moments. For the third- and fourth-order moments, this directly follows from expressions (A8) in Appendix A. (Similar results for velocity moments of higher orders can be obtained using expressions from [4,5].) Expressions for the even velocity moments in terms of the second-order velocity moments derived in such a manner are exactly the same as in a turbulent flow with the Gaussian PDF [6]. That is, for the fourth-order standardized velocity moments, these are expressions (3).

Multiple studies demonstrated that the Gaussian distribution is not an adequate PDF model in turbulent flows (see e.g., [4] and discussion in Ch.6 in [7]). Even in isotropic homogeneous turbulence, PDFs deviate from the Gaussian distribution [8]. In [9], it was shown that in general, the Gaussian distributions should not be expected in homogeneous flows. Nevertheless, the homogeneous shear flow can be considered as an example of a flow with almost Gaussian joint PDFs [10,11].

Increasing the truncation order to $N = 3$ does not change the relations between the higher-order even velocity moments and the second-order ones (see expressions (3) for the fourth-order velocity moments, for example). This means that for purposes of modeling turbulent flows, GCSE should be truncated at minimum $N = 4$.

At $N = 4$, conditions on the standardized velocity moments include those for $l + k = 5$:

$$R^{5,0} = 10R^{3,0},\ R^{4,1} = 6R^{2,1} + 4R^{1,1}R^{3,0},\ R^{3,2} = 6R^{1,1}R^{2,1} + R^{3,0} + 3R^{1,2}$$
$$R^{0,5} = 10R^{0,3},\ R^{1,4} = 6R^{1,2} + 4R^{1,1}R^{0,3},\ R^{2,3} = 6R^{1,1}R^{1,2} + R^{0,3} + 3R^{2,1} \quad (4)$$

and higher [4,5].

Previously, experimental data collected in wall-bounded flows [4,5,12-15] were used to successfully validate the condition $A_{l,k} = 0$ at $N \geq 4$ including expressions (4), with the most complete validation study being conducted in a turbulent boundary layer over a flat plate at the zero pressure gradient [5]. In that study, the expressions for velocity moments through the 11$^{th}$ order obtained by truncating (1) at $N$ = 4, 6, and 8 were validated.

However, even though the previous studies were successful, there were and always will be some expressions left untested because not all moments can be measured and the upper limit for $l + k$ is infinity. Thus, a proof that a truncated GCSE is an accurate representation of a turbulent flow PDF will likely remain incomplete. This issue is common for any PDF model involving series expansions, not only for those based on GCSE.

A more manageable problem of practical importance is validation of expressions (4) as a model for unknown terms describing turbulent diffusion in the Reynolds-averaged Navier-Stokes (RANS) equations for the fourth-order velocity moments [16]. Expressions (4) are particular attractive for modeling purposes as they do not contain unknown coefficients and are applicable to non-Gaussian turbulent flows. After applying expressions (4) to the infinite set of RANS equations, one obtains a fourth-order (or FORANS) statistical closure [16]. Validation of expressions (4) in a wide range of flow geometries and flow conditions is the ultimate goal of our research.

Measuring all moments required to validate (4) is still a complex matter. Direct numerical simulations expand a range of flows and flow conditions, where all expressions (4) can be tested. In the previous study conducted in our group [17], for example, expressions (4) were successfully validated in a two-dimensional zero-pressure-gradient turbulent boundary layer over a flat plate using DNS data from [18-20] at two different Reynolds numbers: $Re_\theta$ = 4101 and 5200, where $Re_\theta$ is based on the free stream velocity and the boundary layer momentum thickness $\theta$. In the current study, a DNS dataset of velocity moments through the fifth order collected in a strained planar turbulent channel flow [21] is used to test validity of expressions (4) in a wall-bounded flow that reverses its direction. (The complete dataset is available in [22]).

Previously, dynamics of the velocity moments through the fourth order and the behavior of the budget terms in their transport equations in this flow were investigated in [21,23]. Here, we will explore in more detail the velocity moment behavior in the near-wall area, where the flow reversal starts. The analysis will also be expanded to include all moments through the fifth order that are relevant to validation of expressions (4) and are useful to understanding effects of the flow reversal on velocity moments of higher orders.

Comparison of the model and DNS profiles of the fifth-order velocity moments revealed a beneficial role of the imposed flow deformation on the profiles agreement in the near-wall flow area known as the buffer zone. Results of investigation of this phenomenon are also presented in the paper.

## 2. DNS data description

In DNS of a turbulent flow in a strained channel [21], simulations were first conducted in a conventional (unstrained) planar fully-developed incompressible turbulent channel flow at $Re_{\tau 0} = 392$, where $Re_{\tau 0}$ is based on the friction velocity $u_{\tau 0}$ and the channel half-height $h$. This corresponds to the Reynolds number $Re_{c0} = U_{c0} h/\nu$ =7,910 based on the centerline mean velocity $U_{c0}$ (hereafter, $\nu$ is the kinematic viscosity of a fluid). The unstrained channel flow data were used then as the initial conditions in simulations of a strained channel flow.

The flow in a strained channel is a time-developing approximation of a spatially-developing adverse-pressure-gradient (APG) turbulent boundary layer (Fig. 1a) as demonstrated in [24]. This is achieved by simultaneously applying uniform irrotational temporal deformations (streamwise compression and wall-normal divergence) to the initial channel flow domain including the walls and by accelerating the walls in the streamwise plane (Fig. 1b). The imposed deformation field varies linearly in space: ${U^*}_i(\vec{x},t) = B_{ij}(t)x_j$, $i = 1,2,3$, where $\vec{x} = (x,y,z)$ and $\overrightarrow{U^*} = (U^*, V^*, 0)$. (The directions $x$, $y$, and z are streamwise, normal-to-the-wall, and spanwise, respectively.) The strain, $B_{ij} = \partial {U^*}_i/\partial x_j$, imposed at $t = 0$ is spatially uniform and constant in time: $\partial B_{ij}/\partial t = 0$. It has two non-zero components: $B_{11} = \partial U^*/\partial x < 0$ (streamwise compression) and $B_{22} = \partial V^*/\partial y > 0$ (wall-normal divergence), which satisfy the condition $B_{11} + B_{22} = 0$.

The wall motion is synchronized with the strain and reproduces the bulk deceleration of APG leading to a reduction in the wall shear stress and eventually, to the flow reversal at the walls. The wall shear stress responds to the difference between the mean centerline velocity $U_c(t)$ and the channel wall velocity $U_w(t)$, both functions of time *t*. The difference is a function of $B_{11}$ and diminishes with time as $U_c(t) - U_w(t) = U_{c0}\exp(B_{11}t)$, where all velocity components are in the streamwise direction and $t$ is time. The wall velocity is specified so that $U_c(t) = U_{c0}\exp(B_{11}t)$ in the reference frame attached to the moving walls.

The procedure described above is equivalent to controlling the pressure gradient, with the flow remaining homogeneous in the stream- and spanwise directions [24]. More information on the flow characteristics and its comparison with a spatially-developing APG boundary layer can be found in [24-26]. Of interest for the current study is that at some instance, the flow in a strained channel undergoes the flow separation near the walls in a sense of the wall shear stress becoming equal to zero at certain time and the flow reversing its direction near the wall as time progresses.

In the simulations [21], the uniform strain magnitude in both directions was set to 31% of the ratio of the initial friction velocity to the initial channel half-width, which gives $|B_{11}| = B_{22} = 0.0118$. DNS data collected at seven non-dimensional times, $t_i' = B_{22}t_i,\ i = 1,..,7$ (Table 1) are used in the current study. The flow separation starts around $t_6'$ =0.675 and continues at $t_7' = 0.772$. The unstrained channel data ($t_0' = 0$) are also used in the study for comparison.

Figure 2a shows dynamics of the flow mean velocity $U$ in the streamwise direction with time, with the separation area being enlarged in Fig. 2b. In the figure, all parameters are taken at a given time moment. Dashed and solid lines show velocity profiles before and after the flow separation, respectively. Labels from 0 to 7 correspond to times $t_i', i = 0, \dots, 7$. At $t_7'$, the separation region is below $y/h = 0.0134$ ($y^+\sim5$). The minimum velocity occurs at $y/h\sim0.005$ ($y^+\sim2$).

**Table 1. Characteristic velocities in DNS of a strained channel flow [18] at different $t_i'$.**

| $\boldsymbol{t'}$ | **0** | **0.02** | **0.1** | **0.191** | **0.285** | **0.365** | **0.675** | **0.772** |
|---|---|---|---|---|---|---|---|---|
| $\boldsymbol{U_c/u_{\tau 0}}$ | 20.137 | 20.493 | 21.948 | 23.764 | 25.773 | 27.848 | 36.975 | 40.511 |
| $\boldsymbol{U_w/u_{\tau 0}}$ | 0 | 0.755 | 3.727 | 7.128 | 10.630 | 13.869 | 26.722 | 31.206 |

## 3. Results and discussion

Since both parameters $u_1$ and $u_2$ used in the definition of $R^{m,n}$ in a turbulent flow can be any component of the turbulent velocity vector $\vec{u} = (u, v, w)$, one can reduce the number of expressions (4) by using the tensor notations:

$$R_{ij}^{5,0} = 10R_{ij}^{3,0},\ R_{ij}^{4,1} = 6R_{ij}^{2,1} + 4R_{ij}^{1,1}R_{ij}^{3,0},\ R_{ij}^{3,2} = 6R_{ij}^{1,1}R_{ij}^{2,1} + R_{ij}^{3,0} + 3R_{ij}^{1,2}, \tag{5}$$

(in Cartesian coordinates), where $R_{ij}^{m,n} = \frac{\langle u_i^m u_j^n \rangle}{\sigma_i^m \sigma_j^n} = \frac{\langle u_i^m u_j^n \rangle}{\langle u_i^2 \rangle^{m/2} \langle u_j^2 \rangle^{n/2}}$, or equivalently, in a form of relations between velocity moments of the fifth-, third-, and second orders:

$$\langle u_i^5 \rangle = 10 \langle u_i^3 \rangle \cdot \langle u_i^2 \rangle ,$$

$$\langle u_i^4 u_j \rangle = 6 \langle u_i^2 u_j \rangle \cdot \langle u_i^2 \rangle + 4 \langle u_i u_j \rangle \cdot \langle u_i^3 \rangle , \tag{6}$$

$$\langle u_i^3 u_j^2 \rangle = 6 \langle u_i u_j \rangle \cdot \langle u_i^2 u_j \rangle + \langle u_i^3 \rangle \cdot \langle u_j^2 \rangle + 3 \langle u_i u_j^2 \rangle \cdot \langle u_i^2 \rangle .$$

Expressions (6) are preferential for validation and modeling purposes as they are free from the potential issue of dividing by zero. For this reason, the following discussion will refer to expressions (6) instead of (5).

Before conducting validation of expressions (6), it is useful to review the time evolution of individual velocity moments in the given flow geometry.

### 3.1 Effects of the flow deformation on velocity moments of different orders

Previous studies [21,23-26] investigated dynamics of the velocity moments through the fourth order in this flow and the behavior of budget terms in their transport equations. In this section, we will expand the analysis to include moments through the fifth order with the purpose of better understanding how the moment order and other moment characteristics affect its response to the imposed flow deformation and what are implications for modeling higher-order velocity moments. Of particular interest is the behavior of velocity moments before and after the flow reversing its direction. For this reason, all plots presented in this section highlight the near-wall area of the flow. Appendix B provides transport equations for the velocity moments appearing in the fourth-order statistical closure prior any modeling in this particular flow geometry. Equations for the velocity moments $\langle u^2 \rangle$ and $\langle u^3 \rangle$ in the streamwise direction and some other velocity moments of the fourth order that do not appear, but may be required when modeling unknown terms, are also available in Appendixes B and C to facilitate the discussion.

It was shown in [24] that response of the Reynolds stresses to the flow deformation is largely driven by the production terms that can be split into two parts: contributions from the flow shear ($\partial U/\partial y$) and from the strain ($B_{11}$). The production by the shear in the transport equation for the $\langle u^2 \rangle$ is larger than the production by the strain at all times except for two small areas: near the flow centerline where the shear tends to zero at any time and at $y^+ < 3$, where velocity moments tend to zero. In the $\langle u^2 \rangle$-equation, the production by the shear is more than 40 times larger than the production by the strain initially. By $t_7'$, contribution of the shear to the production is still four times larger for this velocity moment. The production by the shear is also present in the transport equation for $\langle uv \rangle$, but not in the equation for $\langle v^2 \rangle$, the Reynolds stress in the direction normal to the wall (Appendix B). On the other hand, the strain directly affects the production of $\langle v^2 \rangle$, but not of $\langle uv \rangle$.

The role of the production by the strain in the transport equations of $\langle u^2 \rangle$ and $\langle v^2 \rangle$ in this flow is similar to that in homogeneous turbulence under irrotational plain strain in the directions of flow compression and expansion [27-33]. That is, it contributes to the growth of $\langle u^2 \rangle$ and to the reduction of $\langle v^2 \rangle$ at all times. However, inhomogeneous initial conditions, moving walls, and the shear varying with time change the effects of velocity/pressure-gradient correlations on $\langle v^2 \rangle$ in the strained channel flow to compare with those in homogeneous turbulence. In direct numerical simulations of homogeneous turbulence with various initial conditions [32,34], these correlations were found to be less than the production by the strain. In the strained channel, these correlations are larger than the production by the strain in the budget of $\langle v^2 \rangle$ at any time during the simulation and everywhere in the flow except near the flow centerline. Near the wall, they are the order of magnitude larger than the production by the strain at $t < t_6'$ and about five times larger when the flow undergoes reversal of its direction at later times. As a result, $\langle v^2 \rangle$ initially starts to grow near the wall, to compare with its level in the unstrained channel and opposite to what is observed in homogeneous turbulence.

Increase in $\langle v^2 \rangle$ leads to the initial growth of $\langle uv \rangle$ near the wall, because at $t \leq t_2'$, neither the reduction of the shear, nor changes in velocity/pressure-gradient correlations in the $\langle uv \rangle$-equation are strong enough to overcome the growth of $\langle v^2 \rangle$. However, the effects of the shear start to dominate at $t_3'$ in this Reynolds stress budget pushing the production and the velocity/pressure-correlations below their levels in the unstrained channel. Thus, this moment starts to decrease everywhere, but most rapidly near the wall, where the shear and its effects are the strongest.

At $t_3'$, the Reynolds stress $\langle v^2 \rangle$ is still growing near the wall driven by the velocity/pressure-gradient correlations, but the shear has already started to affect their magnitude, which is at the same level at this time as in the unstrained channel. On the other hand, the strain continues to reduce this moment near the wall and at $t > t_3'$, this Reynolds stress also starts to decrease near the wall following the shear stress. Far from the wall, $\langle v^2 \rangle$ also reduces with time, but much less than near the wall and than the other Reynolds stresses do, because it has no direct contribution from the shear in its budget.

The shear has the strongest effect on $\langle u^2 \rangle$, with the moment growing near the wall only at $t_1'$ due to the production by the strain. Changes in the production by the shear and in the velocity/pressure-gradient correlations are minor for this Reynolds stress at this time, but these changes contribute in the reduction of its magnitude. That is, the growth of $\langle uv \rangle$ is insufficient to overcome the shear effects even at that time. At $t_2'$, the level of the production by the shear in the budget for this moment is below than in the unstrained channel, but the velocity/pressure-gradient correlations still grow, which further reduces this moment. At $t_3'$, the velocity/pressure-gradient correlations also drop below their level in the unstrained channel. The production by the shear remains the dominant term in the budget of this moment in the remaining time of the simulation except for the areas near the flow centerline and at $y^+ < 10$. During the flow reversal at $t \geq t_6'$, this term and the velocity/pressure-gradient correlations determine the profile of $\langle u^2 \rangle$ outside the wall area, where all terms are small.

To sum, with respect to dynamics of the Reynolds stresses near the wall caused by the imposed flow deformation, one can recognize the initial transitional period, $t \leq t_3'$, when the production by the shear and the velocity/pressure-gradient correlations may exceed their levels in the unstrained channel flow, which lead to temporal growth of Reynolds stresses associated with the velocity fluctuation in the direction normal to the wall. The production by the strain also causes the short-lived growth of the Reynolds stress in the streamwise direction. At $t > t_3'$, the flow deceleration results in the production by the shear and the velocity/pressure-gradient correlations falling below their levels in the unstrained channel flow, and all three Reynolds stresses decrease everywhere in the flow as time progresses except near the flow centerline, where the shear tends to zero. Additional information on the flow deformation effects on all terms in the budgets of Reynolds stresses in the strained channel can be found in [21,23-26].

Figures 3, 4, and 5 illustrate transformation of the profiles of Reynolds stresses and higher-order velocity moments with time under the imposed flow deformation. In the figures, the shown

moment profiles are those in the unstrained flow ($t_0'$), in pre-separated and separated flows ($t_5'$ - $t_7'$), and at the time when a velocity moment reaches its maximum absolute value (this time varies for different moments).

In the budgets of higher-order velocity moments (Appendices B and C), the production terms have the structure similar to those in the Reynolds stress transport equations, with both the shear and the strain directly affecting velocity moments $\langle u^m \rangle$: $-m(\langle u^m \rangle B_{11} + \langle u^{m-1} v \rangle \partial U/\partial y)$, and with only the strain appearing in the $\langle v^m \rangle$-equations as the term $m\langle v^m \rangle B_{11}$. In the equations for all cross-correlations but $\langle u^2 v^2 \rangle$, there are contributions from both the strain and the shear in the velocity moment production. In the $\langle u^2 v^2 \rangle$-equation (not shown here), there is only production by the shear, similar to the $\langle uv \rangle$-equation.

Similarity in the transport equations of velocity moments of different orders is reflected in similar responses of the moments to the flow deformation. Specifically, all velocity moments $\langle u^m \rangle$, $m = 2, \ldots, 5$, in the streamwise direction achieve their maximum values at $t_1'$ (Fig. 3) regardless the moment order, *m*, but the difference in the profiles at $t_1'$ and at $t_0'$ is minor for all *m*. On the other hand, all these moments experience strong reduction in their magnitudes as time progresses: the higher the moment order, the stronger suppression.

The velocity moments $\langle v^m \rangle$ in the direction normal to the channel wall reach their maximum later in time than $\langle u^m \rangle$, with the odd moments $\langle v^3 \rangle$ and $\langle v^5 \rangle$ achieving their maximum values at $t_2'$ and the even moments $\langle v^2 \rangle$ and $\langle v^4 \rangle$ at $t_3'$ (Fig. 4). With time increasing, the magnitude of all moments $\langle v^m \rangle$ vary less than that of $\langle u^m \rangle$, with the odd moments being affected stronger by the flow deformation than the even ones. When comparing only odd or only even moments, the effects are stronger for higher order moments.

The ratio between the maximum absolute values of a velocity moment at $t'_7$ and $t'_0$:

$$r = \frac{max\left|\langle u_i^m u_j^n \rangle\right|(t'_0)}{max\left|\langle u_i^m u_j^n \rangle\right|(t'_7)}, \tag{7}$$

is used in the current study to quantify and to compare responsiveness of different velocity moments to the flow deformation (Table 2). Turbulent models are expected to accurately reproduce this parameter for individual moments.

Overall, changes in all $\langle v^m \rangle$ with the time are less dramatic than in all $\langle u^m \rangle$, which is due to the lack of direct contribution from the shear in the production of these moments and the role of the velocity/pressure-gradient correlations present in dynamics of these moments. The difference in behavior of the odd and even moments can be linked to the fact that the odd moments of the $m$ order are smaller than the even moments of the $(m-1)$ and $(m+1)$ orders. As a result, a role of the turbulent diffusion terms is different in the budgets of the odd and even moments: the major contributor to the budget of $\langle v^3 \rangle$ and the secondary one to the budgets of $\langle v^2 \rangle$ and $\langle v^4 \rangle$. In addition, new terms like $\langle v^{m-1} \rangle \frac{\partial \langle v^2 \rangle}{\partial y}$ ("production by turbulence") appear in the transport equations of the velocity moments of the orders $m > 2$, and they are larger in the budgets of the odd moments tending to reduce them. On the other hand, the even moments have strong support from the velocity/pressure-gradient correlations at any time, whereas their contribution to the odd moments reduces strongly as time progresses. Other aspects of dynamics of the budget terms in the transport equations of higher-order velocity moments under the flow deformation are discussed in [21, 23].

When considering cross-correlations $\langle u^m v^n \rangle$, where $m + n \leq 5$, all those with $m > n$ reach their maximum at $t_1'$. The others have maximum values at $t_2'$, similar to the odd moments $\langle v^3 \rangle$ and $\langle v^5 \rangle$.

As time progresses, all moments $\langle u^m v^n \rangle$ are suppressed in a less degree than $\langle u^{m+n} \rangle$, but more so than $\langle v^{m+n} \rangle$ at the same moment order, with the even moments ($m + n = 2{,}4$) being affected less than the odd ones (Table 2). The moments with $m > n$ are suppressed more than other cross-correlations.

**Table 2. Ratio *r* (Eq. (7)) for velocity moments in Figs. 3-6.**

| | | | | | | | | |
|---|---|---|---|---|---|---|---|---|
| $\langle u^5 \rangle$ | $\langle u^4 \rangle$ | $\langle u^3 \rangle$ | $\langle u^2 \rangle$ | $\langle uv \rangle$ | $\langle v^2 \rangle$ | $\langle v^3 \rangle$ | $\langle v^4 \rangle$ | $\langle v^5 \rangle$ |
| 15.41 | 7.20 | 6.48 | 3.02 | 1.28 | 1.03 | 1.39 | 1.15 | 1.71 |
| | $\langle u^4 v \rangle$ | $\langle u^3 v \rangle$ | $\langle u^2 v \rangle$ | $\langle u^2 v^2 \rangle$ | $\langle uv^2 \rangle$ | $\langle uv^3 \rangle$ | $\langle uv^4 \rangle$ | |
| | 8.73 | 2.95 | 3.69 | 1.69 | 1.6 | 1.45 | 1.95 | |
| | | $\langle u^3 v^2 \rangle$ | | | | $\langle u^2 v^3 \rangle$ | | |
| | | 4.07 | | | | 2.76 | | |

Comparing all moments, the moment with the most reduced maximum value under the imposed flow deformation is $\langle u^5 \rangle$ and the moment with the least affected maximum value is $\langle v^2 \rangle$ closely followed by $\langle v^4 \rangle$.

In fact, of all considered moments only these two, $\langle v^2 \rangle$ and $\langle v^4 \rangle$, have their maximum values reduced to compare with those in the unstrained channel at the same time when the flow separation occurs, that is at $t_6'$ and not sooner than that. Since the transport equations for these two moments (Appendix B) do not explicitly contain terms with $\partial U / \partial y$, this is an indication that models for unknown terms in these equations have to provide a link between these moments and the mean velocity gradient in the direction normal to the wall.

As the next step of the analysis, we compare the largest and the smallest moments of the same order in an unstrained channel (at $t_0'$) and when the flow undergoes separation (at $t_7'$). This is also important information characterizing the flow turbulence and the flow response to the flow deformation, which turbulence models are expected to accurately reproduce. The largest moment at any order $(m+n)$ and considered times is the one in the streamwise direction, $\langle u^{m+n}\rangle$. Thus, the ratio of the maximum absolute values of $\langle u^{m+n}\rangle$ and of the smallest moments at a given velocity moment order $(m+n)$:

$$r1(t_i'; m+n) = \frac{max|\langle u^{m+n}\rangle|}{min\,(max|\langle u^m v^n\rangle|)} \quad (8)$$

is used to evaluate this effect. In (8), $i = 0, 7$ and $m+n = 2, \dots, 5$. The smallest moment varies depending on the moment order and time. At $m+n < 5$, the moments $\langle uv\rangle, \langle v^3\rangle, \langle uv^3\rangle$ are the smallest at both times. At $m+n = 5$, the smallest moment at $t_0'$ is $\langle v^5\rangle$, but $\langle uv^4\rangle$ at $t_7'$. Results are presented in Table 3:

**Table 3. Ratio *r*1 (Eq. (8)).**

| Moment order, $m+n$ | 2 | 3 | 4 | 5 |
|---|---|---|---|---|
| $t_0'$ | 8.87 | 20.40 | 52.10 | 84.75 |
| $t_7'$ | 3.77 | 4.38 | 10.48 | 10.35 |

The table shows that the largest differences between the moments of the same orders occur in the unstrained channel, with the ratio $r1$ growing with the moment order. It is on the order of magnitude less at $m + n = 2$ to compare with $m + n = 5$.

The imposed flow deformation tends to reduce differences between the moments of the same order and more so for the moments with higher values of $(m + n)$. As a result, the $r1$-value is still growing with the moment order, but more slowly than in the unstrained channel. At $m + n$ = 5, it is of O(10) to compare with O($10^2$) in the unstrained channel.

Based on the above data analysis, challenges of modeling fifth-order velocity moments in terms of lower-order ones become more apparent as different velocity moments respond differently to the imposed flow deformation with the following complex evolution of the moment profiles in time. However, the flow deformation (at least during the conducted simulation time) seems to have favorable effects on the moment dynamics from the modeling perspective by smoothing differences rather than exaggerating them. Complexity of the moment transformation under the flow deformation makes it unlikely to model relations between different moments in intuitive empirical fashion. This makes expressions (6) even more valuable as they are rigorously derived and do not contain unknown model parameters to vary. This is of course, if these expressions prove to be successful in this challenging flow. The following sub-section describes results of their validation.

### 3.2 The fifth-order moments modeling

A quick look at Figs. 3-6 encourages the idea of modeling higher-order moments in terms of lower-order ones based on the obvious similarity of some moment profiles, their dynamics with time, and times when the moments reach their maximum. Compare, for example, $\langle u^5 \rangle$ and $\langle u^3 \rangle$, $\langle v^5 \rangle$ and $\langle v^3 \rangle$.

In the strained channel flow that remains homogeneous in the stream- and spanwise directions under the imposed flow deformation [24], there are only six equations in the FORANS set for velocity moments to solve (Appendix B). In these equations, there are only two fifth-order moments: $\langle uv^4 \rangle$ and $\langle v^5 \rangle$, to model. This is, however, before modeling unknown terms.

Model expressions (6) for $\langle uv^4 \rangle$ and $\langle v^5 \rangle$: $\langle uv^4 \rangle = 6\langle v^2 \rangle \cdot \langle uv^2 \rangle + 4\langle v^3 \rangle \cdot \langle uv \rangle$ and $\langle v^5 \rangle = 10\langle v^2 \rangle \cdot \langle v^3 \rangle$, do not bring new unknown parameters (and transport equations for them) into the set of equations (B1)-(B6). This is an additional benefit of the considered modeling approach.

On the other hand, models for velocity/pressure-gradient correlations and for components of the turbulent kinetic energy dissipation tensors are likely to add new parameters to the original set of equations (B1)-(B6) governing this flow. As a result, one may expect to have more fifth-order velocity moments to model. For this reason, we validate in this study expressions (6) for all fifth-order velocity moments that appear in the RANS equations for the fourth-order velocity moments in an incompressible planar turbulent flow in the (*x*-*y*) plane (Appendix C).

Figure 7 compares profiles of the fifth-order moments $\langle u^m v^n \rangle$, $m + n = 5$, obtained from DNS data [21,22] (dashed lines) with the model profiles $\langle u^m v^n \rangle^{(M)}$ of the same moments obtained from expressions (6) using DNS data for the third- and second-order moments (solid lines). Comparison is made at different times to demonstrate that dynamics of the model profiles of the fifth-order moments is consistent with that of their DNS profiles. The logarithmic vertical scale is chosen for the moments $\langle u^m v^n \rangle$ with $m = 4$ and 5, to better resolve the area close to the wall, where the magnitudes of these moments reach their maximum.

Overall, the agreement between the DNS and model profiles is very good for all moments everywhere in the flow. Dynamics of the model profiles with time is the same as that of the DNS profiles, with the agreement between the DNS and model profiles improving with time (as expected and discussed in the previous sub-section). In a presence of the flow separation, the agreement is excellent. Locations of the maximum and minimum values as well as those of zeros are well reproduced at every time. These results are very encouraging from the modeling perspective.

However, there is a discrepancy between the model and DNS profiles, which appears to be sensitive to the turbulent velocity fluctuation in the streamwise direction, that is, the discrepancy grows with $m$ in $\langle u^m v^n \rangle$. The observed tendencies for this parameter are illustrated in Fig. 8 for $\langle u^5 \rangle$ and $\langle v^5 \rangle$ at $t_0'$ and $t_7'$ using the following ratio based on the $L_\infty$-norms:

$$\Delta = \frac{\left| \langle u^m v^n \rangle^{(M)} - \langle u^m v^n \rangle \right|}{\left| \langle u^m v^n \rangle \right|_\infty} . \tag{9}$$

This ratio was found to be more informative than others (not reported here) as it allows avoiding division by small numbers when a moment changes its sign (and in this flow, all moments do). Results for the other four moments (not shown here) fall between those in the figures. Profiles of velocity moments $\langle u^5 \rangle$ and $\langle v^5 \rangle$ are also shown in Fig. 8 at both times by dashed lines for comparison.

For $\langle v^5 \rangle$, the discrepancy between the DNS and model profiles peaks near or at the locations of the velocity moment zero values at both times and can be explained by numerical errors associated with small numbers. The magnitude of this parameter does not change significantly with time (Figs. 8c and 8d, also Table 4).

For $\langle u^5 \rangle$, on the other hand, the discrepancy between the DNS and model profiles tends to peak around the locations of the extremes in this moment profiles and more so at earlier times when this moment magnitudes are the largest. At $t_0{}'$, for example, the Δ-profile has two prominent extremes (Fig. 8a) located at $y^+ = 8.8$ and 20. The $\langle u^5 \rangle$-profile also has two extremes located at $y^+ = 7.5$ and 23 (see also Table 4 where locations of their maximum values are compared). As time progresses and this moment magnitude reduces, a connection between locations of the extremes in the profiles of Δ and $\langle u^5 \rangle$ remains until the flow separates (Fig. 8b and Table 4). Still, even in the separated flow, the location of max Δ at $y^+ = 185$ is not that far from the location of max$|\langle u^5 \rangle|$ at $y^+ = 235$.

Notice that before the flow separates (at $t < t_6{}'$), the extreme values of Δ and that of $\langle u^5 \rangle$ are located within the buffer zone at $y^+ \lesssim 20$.

The magnitude of Δ can also be linked to that of $\langle u^5 \rangle$. As discussed in the previous sub-section, $\langle u^5 \rangle$ is the moment suppressed the most by the imposed flow deformation. In the separated flow, the $\langle u^5 \rangle$ -magnitude reduces at least ten times to compare with its level at $t_0{}'$. The same trend is observed for the Δ-values (Fig. 8b and Table 5).

**Table 4. Non-dimensional locations $y/h$ of max $\Delta$ and max $|\langle u^5\rangle|$ at different times.**

| | $t'_0$ | $t'_1$ | $t'_2$ | $t'_3$ | $t'_5$ | $t'_6$ | $t'_7$ |
|---|---|---|---|---|---|---|---|
| $y/h(\max\Delta)$ | 0.05 | 0.05 | 0.05 | 0.02 | 0.02 | 0.6 | 0.5 |
| $y/h(\max\langle u^5\rangle)$ | 0.06 | 0.06 | 0.02 | 0.02 | 0.02 | 0.55 | 0.64 |

**Table 5. Max $\Delta$ for different moments.**

| | $\langle u^5\rangle$ | $\langle u^4v\rangle$ | $\langle u^3v^2\rangle$ | $\langle u^2v^3\rangle$ | $\langle uv^4\rangle$ | $\langle v^5\rangle$ |
|---|---|---|---|---|---|---|
| $t'_0$ | 0.68 | 0.57 | 0.43 | 0.25 | 0.12 | 0.06 |
| $t'_1$ | 0.67 | 0.56 | 0.42 | 0.25 | 0.12 | 0.06 |
| $t'_2$ | 0.62 | 0.54 | 0.43 | 0.27 | 0.14 | 0.06 |
| $t'_3$ | 0.32 | 0.49 | 0.41 | 0.24 | 0.17 | 0.06 |
| $t'_5$ | 0.26 | 0.24 | 0.18 | 0.19 | 0.18 | 0.11 |
| $t'_6$ | 0.12 | 0.17 | 0.14 | 0.16 | 0.14 | 0.12 |
| $t'_7$ | 0.13 | 0.17 | 0.15 | 0.17 | 0.15 | 0.11 |

Finally, comparing the maximum values of $\Delta$ for all fifth-order moments at different times, one can notice that its largest value happens for $\langle u^5\rangle$ in the unstrained channel (Table 5). Also, all moments $\langle u^m v^n\rangle$ with $m > n$ have larger maximum $\Delta$-values than those with $m < n$ prior the flow separation. The least affected by the flow evolution is the maximum $\Delta$-value for $\langle v^5\rangle$.

To summarize, the discrepancy between the DNS and model profiles can be linked to the magnitude of velocity fluctuations in the streamwise direction and to a specific location within the flow: the buffer zone. The difference is the largest in the unstrained channel, where the velocity moment $\langle u^5\rangle$ is the largest. The imposed flow deformation reduces $\langle u^5\rangle$, which leads to improvement of the agreement between the DNS and model profiles for this moment. Similar tendencies are observed for other moments $\langle u^m v^n\rangle$ with $m > n$. The discrepancy between the DNS and model profiles for $\langle v^5\rangle$ and for other moments $\langle u^m v^n\rangle$ with $m < n$ remains small and changes little with time.

For this reason, the following sub-section further scrutinizes DNS data in the unstrained channel to better understand how velocity fluctuations in the streamwise direction contribute into the observed discrepancy between the model and DNS profiles.

**3.3 Analysis of discrepancy between the model and DNS profiles**

The discussion in the previous sub-section points towards the turbulence production by the mean velocity gradient $\partial U/\partial y$ as a major culprit of the discrepancy between the DNS and model profiles obtained using (6). Indeed, these are the only terms that are absent in the transport equations for moments $\langle v^n \rangle$. The contribution of these terms grows with $m$ in the transport equations for moments $\langle u^m v^n \rangle$: $(-m\langle u^{m-1} v^{n+1} \rangle \partial U/\partial y)$ (see Appendices B and C and also [35]). Contribution of these terms also reduces with the imposed flow deformation.

To better understand how the production by the shear terms may contribute in the discrepancy between the DNS and model profiles obtained from (5), the following discussion will be limited to the moment $\langle u^5 \rangle$ in the unstrained channel, where these effects are pronounced the most.

In the model for $\langle u^5 \rangle$ (Eq. (6)), two lower-order moments are involved: $\langle u^2 \rangle$ and $\langle u^3 \rangle$. Figure 9 shows the turbulence production terms by $\partial U/\partial y$ (absolute values) in the transport equations for the three moments (dashed and dotted lines) as well as the Δ- profile for $\langle u^5 \rangle$ (solid line). The production terms are not to scale as the purpose of this plot is to demonstrate a connection between the locations of the extreme values in the Δ-profile and those in the profiles of production terms.

There are two largest extremes in the Δ-profile located at $y^+ = 8.8$ and 20.8. The production term for $\langle u^2 \rangle$ (dotted line in Fig. 9) has only one maximum located at $y^+ = 15.1$ between the two extremes in the Δ-profile. Thus, this is not the leading term contributing to Δ. On the other hand, both profiles for the productions terms in the transport equations for the odd moments $\langle u^3 \rangle$ and $\langle u^5 \rangle$ (shown by the dashed lines in Fig. 9) have two largest extremes closely aligned between the profiles and with the extremes in the Δ-profile. Specifically, in the $\langle u^3 \rangle$-profile, their locations are $y^+ =$7.5 and 20.1 and in the $\langle u^3 \rangle$-profile, the extremes are located at $y^+ = 8.8$ and 20.8.

Observation that the odd moments contribute more in the discrepancy between the DNS and model profiles for $\langle u^5 \rangle$ is consistent with our previous results [35], where RANS-DNS simulations were used to demonstrate that the errors in the DNS $\langle u^3 \rangle$-profile are significantly larger than those in the profiles of even moments $\langle u^2 \rangle$ and $\langle u^4 \rangle$. Results were obtained with the same data [21,23] as used in the current study. No comments on the accuracy of $\langle u^5 \rangle$ can currently be made using the RANS-DNS simulations due to the lack of the complete dataset for the budget terms in the transport equation for this moment.

Another interesting feature to notice in Fig. 9 is that alignment of the extremes in the profiles of $\langle u^3 \rangle$, $\langle u^5 \rangle$, and Δ is “reverse”: the smallest of the two extremes in the Δ-profile corresponds to the maximum production values. This may be an indication that the odd moments are “under-produced” in this flow area. This can happen, for example, if the collecting time of raw DNS data was not sufficient for the profiles of odd moments to statistically converge or the grid resolution may not be sufficient to resolve rapid variations in the moment values, or both.

Insufficient statistical convergence of DNS data is habitually brought into discussion whenever there is an issue associated with such data. However, convergence of statistics including high-order moments was tested, discussed at length and found adequate in [21]. In addition, the same DNS in the unstrained channel were run about ten times longer in our separate study [36]. In [36], RANS-DNS simulations were used to analyze the data convergence (a different method than used in [21]). They confirmed that statistical convergence of the second-order moments and of the budget terms in their transport equations occurs early in the simulations, close to the simulation time in [21].

On the other hand, that study also revealed a systematic error present in the second-order statistics. It can be seen, for example, in the balance error profiles in the budgets of the Reynolds stresses that converged to non-zero values at the end of simulations, with the maximum values being located in the flow buffer zone. As an example, Fig. 10a shows the non-dimensional balance errors $Err_{xx}$ in the $\langle u^2 \rangle$-budget obtained at the end of simulations at $t = 1.22 \cdot 10^5$ (compare with the simulation time of $t = 5.06 \cdot 10^4$ in [21]). The simulation times in [36] and [21] correspond to 1000 and 159 instantaneous flow field realizations, respectively. In this and following figures, the balance errors and other budget terms in the Reynolds-stress transport equations are normalized by $u_{\tau 0}^4 / \nu$ [23].

Higher-order statistics were not collected in [36], but as has already been mentioned above, their statistical convergence was found acceptable in [21] using standard approaches to the statistical error estimation. As for the systematic error discovered in [36] in the second-order statistics, it is unlikely to expect higher-order statistics to be free from this error. Moreover, one would expect such error to be larger in odd moments as they are sensitive to the direction of velocity fluctuations in contrast to even moments.

Indeed, using data from [21], Fig. 10b confirms that the balance errors in the transport equation for $\langle u^3 \rangle$ are significantly larger than those in the $\langle u^2 \rangle$ −equation. They are the largest in the buffer zone, where the systematic error is expected to be the largest, and their extreme values are closely aligned with those of the production term in the $\langle u^3 \rangle$-equation: $y^+ =$6.3 and 18.8 vs. $y^+ =$7.5 and 20.1 in the production profile. Similar to what was observed in Fig. 9, the highest peak from the two extreme values in the production term corresponds to the lower of the two extremes in the balance error profile.

To compare, there is no obvious relation between the balance errors in the transport equation for $\langle u^2 \rangle$ and the production term in that equation also shown in the figure. Profiles for the terms in the $\langle u^3 \rangle$-equation shown in Fig. 10b and the following figures are normalized by $u_{\tau 0}^5/\nu$ [23].

To summarize the discussion above, the largest discrepancy between the DNS and model profiles (6) observed in the flow buffer zone can be linked to the production terms in the transport equations of the odd moments.

In this flow geometry, the buffer zone is the area where the considered velocity moments and the production terms in their transport equations reach their maximum values. This is also the region where the balance errors in the DNS budgets of velocity moments are the largest, with the extreme values in their profiles being aligned with those in the production terms in the transport equations of the odd moments.

Since the statistical errors in DNS data cannot alone explain errors in DNS data in the buffer zone, this leaves the grid resolution as a possible source of the systematic error in DNS data [36]. However, the grid used in [21] was generated following the standard approach in which the Kolmogorov turbulence length scale [37]

$$\eta = (\nu^3/\varepsilon)^{1/4}, \tag{10}$$

is used as a reference to compare the grid resolution with. In (10), $\varepsilon$ is the scalar dissipation, $\varepsilon = 1/2\sum_{i=1,..,3}\varepsilon_{ii}$, where $\varepsilon_{ii}$ are the dissipation tensor components in the transport equations for Reynolds stresses $\langle u_i^2\rangle$.

The Kolmogorov length scale is considered to be the smallest turbulence scale to resolve based on the theory of locally isotropic turbulence (see [37] and also [38,39], where history of the theory development is discussed). The theory assumes, in particularly, a sufficiently large Reynolds number and small velocity fluctuations. Applicability of such assumptions to the buffer zone in wall-bounded flows, where the mean velocity gradients are high, local Reynolds numbers are small, and turbulence is highly anisotropic (Figs. 3-6) is questionable. In fact, it was explicitly stated in [37] that "the hypothesis of local isotropy is realized with good approximation in sufficiently small domains *G* of the four-dimensional space $(x_1, x_2, x_3. t)$ **not lying near the boundary of the flow or its other singularities**."

As an example, Fig. 11a compares local velocity scales defined as $\sqrt[n]{|\langle u^n\rangle|}$ with the local mean velocity. Ratios $\sqrt[n]{|\langle u^n\rangle|}/U$ are shown by blue and red lines for even and odd moments, respectively. Also shown are the balance errors in the transport equations for $\langle u^2\rangle$ and $\langle u^3\rangle$ (black lines). Solid color lines correspond to the lower-order moments ($n \le 3$) and dashed color lines are for the moments with $n = 4$ and 5. Notations for the balance errors are the same as in Fig. 10b.

As the figure demonstrates, the ratio $\sqrt[n]{|\langle u^n\rangle|}/U$ is not small for all considered $n$ and in the large flow area, but particularly, within the buffer zone at $y/h < 0.13$ ($y^+ < 50$). Moreover, the ratio grows with $n$, whereas the velocity scale extracted from $\sqrt[n]{|\langle u^n \rangle|}$ should be independent from $n$ in accurate simulations.

The two odd moments have a dip in their profiles at the location where both of them change sign. The ratio value should be close to zero in this location, but instead, it is O(0.1) for both moments, another evidence of unresolved values of the odd moments. The dip is located close to the locations of the extreme values in the profiles of the balance errors and of the production by $\partial U/\partial y$ in the $\langle u^3 \rangle$–budget. Notice that there is no apparent relation between the ratio $\sqrt[n]{|\langle u^n \rangle|}/U$ and the balance errors in the flow area of $y^+ < 2$, but this is the region where all velocity moments diminish (Figs. 3-6) and so do the balance errors in the moment budgets.

Additional evidence that the Kolmogorov lengthscale may not be an adequate reference for determining the grid resolution in the buffer zone can be found in Fig. 12. In the figure, variation of the non-dimensional lengthscale $L_k^* = \eta/h$ with $y$ is shown by the dash-dotted line. The lengthscale remains almost unchanged through this area, whereas the balance errors are at their largest values. The non-dimensional grid cell size, $L_{14}^* = \Delta y/h$, used in [21] is also shown in the figure by the long-dashed line. This parameter remains below $L_k^*$ at $y^+ < 14$ ($y/h < 0.035$), but after that, exceeds it everywhere in the flow. Yet, the balance errors in the budgets of both moments $\langle u^2 \rangle$ and $\langle u^3 \rangle$ start to diminish regardless of the increase in $L_{14}^*$.

The main conclusion from the results shown in Fig. 12 is that the balance errors start to be noticeable when $L_{14}^*$ reaches a certain threshold of ~0.00134 at $y^+ \sim 1.6$ ($y/h \sim 0.004$). Outside the buffer zone, the simulation results are much less sensitive to the grid cell size, although fluctuations present in the balance errors of the $\langle u^3 \rangle$ - budget may be an indication that $L_{14}^*$ is rather large for the odd moments.

### 3.4 Comparison with the previous studies

Previously, we compared the DNS and model profiles obtained from (5) in a different flow geometry: a planar turbulent boundary layer over a flat plate under zero pressure gradient [17]. Raw DNS data from [20] at two different Reynolds numbers, $Re_\theta = 4101$ and 5200, were used to extract velocity moments. The different code and the grid were used in DNS in [20] to compare with those in [21]. Yet, there is similarity in the results obtained in the current work and in [17]: the model profiles for $\langle u^m v^n \rangle$ obtained in [17] were higher than DNS profiles for the corresponding velocity moments, with the discrepancy between the DNS and model profiles being the largest at the locations of the velocity moment extreme values. The discrepancy increased with *m* (Fig. 7 in [17]), with the largest discrepancy being between the DNS and model profiles of $\langle u^5 \rangle$. The agreement between the DNS and model profiles of $\langle v^5 \rangle$ were again excellent everywhere in the flow. For $\langle u^5 \rangle$, the extreme values and the largest deviation between the DNS and model profiles were found to be within the flow buffer zone ($y^+ < 50$).

Not much can be said about quality of the grid used in [20] though, as the budget terms are only available for the second-order moments in this flow. Data at $Re_\theta = 5200$ are used in the discussion below.

The non-dimensional grid cell size $L_{13}^* = \Delta y / \delta_{99}$ is found to be smaller than the Kolmogorov lengthscale everywhere in this flow (Fig. 12b), but again, the balance errors in the $\langle u^2 \rangle$-budget are in no particular correlation with the grid cell size growth through the flow. The error absolute values start to increase right from the wall in contrast to what is seen in Fig. 12a with data from [17]. Interestingly, the Kolmogorov lengthscale is not a smooth function in simulations [20], as both the scalar dissipation profile and the dissipation profile in the $\langle u^2 \rangle$- budget change their sign on several occasions in the near wall area. Whether this is a physical phenomenon or a simulation artifact remains unclear as the balance errors are the largest in this area. (Notice that the Kolmogorov lengthscale is smooth when calculated using data from [21,23] (see Fig. 12a).) Extreme values of the balance errors are confined within $y^+ < 80$. The production term has its maximum at $y^+$=10.4 and its magnitude reduces ten times within $y^+ < 80$. The same occurs in the budget of $\langle u^2 \rangle$ from [21,23]. There is no particular alignment between the locations of the maximum production value and the balance error extreme values, another similarity between the data from [20] and [21,23].

Thus, although a relation between the balance errors and the production term in the $\langle u^2 \rangle$-budget can be detected, the relation is not that obvious to compare with what we saw when analyzing the budget of $\langle u^3 \rangle$ using data from [21,23]. That is, collecting budgets for odd moments should not be treated as optional, but necessary for the analysis of DNS accuracy. This is because odd moments are sensitive to the sign of velocity fluctuations in the given flow direction and therefore, reveal more information about the flow structure and development than even moments.

Lastly, the ratio $\sqrt[n]{|\langle u^n \rangle|}/U$ for $n$ = 2 and 3 is presented for this flow (Fig. 11b). Once again, the ratio is not small in a large portion of the flow and particularly, near the wall demonstrating that the theory of locally isotropic turbulence is inapplicable in this area. The dip in the ratio value at $n$ = 3 is resolved better than with the data from [21,23]. However, this ratio also shows that the moment $\langle u^3 \rangle$ from [20] is less accurate than in [21,23], because the ratios at $n$ = 2 and 3 deviate from each other farther everywhere in the flow in Fig. 11b than in Fig. 11a. These results are in agreement with our previous observations [35], where the errors in both datasets were analyzed using the RANS-DNS framework.

### 3.5 New length scales for high mean velocity gradient flow areas

The analysis of errors conducted in the previous section revealed a connection between the turbulence production by the mean velocity gradient in the flow buffer zone, where these gradients are high, and the balance errors in the budgets of particularly odd velocity moments in this flow area. Insufficient grid resolution was identified as the main contributor to the balance errors. This section provides a discussion on possible alternatives to the Kolmogorov length scale in this flow region (and potentially, in others with high mean velocity gradients) that will help to improve the grid quality when collecting high-order statistics and particularly of odd orders. Such length scales are expected to be smaller than the Kolmogorov scale in the areas with high mean velocity gradients. The dimensional analysis is used to obtain the scales.

All considered scales involve viscosity, as relevant effects are high in the buffer zone. Other parameters considered in determining alternative length scales include the production terms in the transport equations of velocity moments $\langle u^n \rangle$, the mean velocity gradient $\partial U/\partial y$, mean velocity $U$, it's the second derivative $\partial^2 U/\partial y^2$ and velocity moments $\langle u^n \rangle$ as the largest ones. From all these options, results for those scales that are based on $\langle u^n \rangle$ and the productions terms are discussed here, as other lengthscales turned out to be larger than the Kolmogorov length scale everywhere in the flow.

A lengthscale based on the production terms for arbitrary order $n$ of the corresponding velocity moment can be defined as:

$$L_{pn} = \left(\frac{\nu^{n+1}}{|P_n|}\right)^{\frac{1}{n+2}} , \tag{11}$$

where $P_n$ is the production term by $\partial U/\partial y$ in the transport equation for $\langle u^n \rangle$.

DNS data are often presented in a non-dimensional form. In [23], for example, viscosity, the channel half-height and the friction velocity in an unstrained channel were used for this purpose, with all budget terms being normalized by $u_{\tau 0}^{n+2}/\nu$. For the production terms, it gives, $P_n = P_n^* \cdot u_{\tau 0}^{n+2}/\nu$, where $P_n^*$ is the non-dimensional production term from the DNS dataset for a corresponding velocity moment. Then, the non-dimensional length scale $L^*{}_{pn} = L_{pn}/h$ can be derived from (11) as

$$L^*{}_{pn} = \frac{1}{Re_{\tau 0}} \left(\frac{1}{|P_n^*|}\right)^{\frac{1}{n+2}} . \tag{12}$$

Notice that the scale has the limit at $n \to \infty$:

$$L^*{}_{inf} = \frac{1}{Re_{\tau 0}} . \tag{13}$$

This is a useful limit when one concerns with collecting statistics of different orders. It is also easier to find an estimate for (13) prior conducting expensive simulations. Moreover, as previously discussed, the production terms are likely underpredicted in DNS in the flow areas with high-velocity gradients due to insufficient grid resolution. Expression (13) will not have this issue.

Since errors in the production terms affect velocity moments, and because an estimate for a velocity moment is easier to obtain than that for the production terms prior simulations, a lengthscale based on viscosity and a velocity moment of the *n*-th order was also derived:

$$L_{un} = \frac{\nu}{(|\langle u^n \rangle|)^{1/n}}. \tag{14}$$

Using relations between dimensional and non-dimensional parameters: $(\langle u^n \rangle)^{\frac{1}{n}} = u_{\tau 0}(\langle u^{*n} \rangle)^{\frac{1}{n}}$, where $u^*$ is a non-dimensional velocity fluctuation, and $L^*{}_{un} = L_{un}/h$, expression (14) can be re-written as:

$$L^*{}_{un} = \frac{1}{Re_{\tau 0}(|\langle u'^n \rangle|)^{1/n}}, \tag{15}$$

which at $n \to \infty$ becomes again expression (13).

Whereas the exact physical process associated with the lengthscale given by expression (13) has yet to be understood, this process is likely to be relevant to formation and dynamics of small-scaled coherent structures in the buffer zone investigated in many previous works, with the main current findings being summarized in [40]. In particular, there is a consensus that dynamics of these coherent structures inside the buffer zone is driven by viscosity and the mean shear, this dynamics is different and independent from the rest of the flow, but common with that observed in statistically stationary homogeneous shear turbulence. That is, the phenomenon is due to the mean shear, not the wall. Some questions remain though [40]. DNS of statistically stationary homogeneous shear turbulence with the grid refined to resolve processes at the lengthscale analogous to (13) in a wall-less flow will be beneficial for answering those questions. Collecting data for better understanding a possible relation between one-point high-order velocity moments and coherent structures will also be beneficial for improving turbulence modeling.

Figure 13 compares variation of five lengths scales: $L^*{}_{inf}$, $L^*{}_{p2}$, $L^*{}_{p3}$, $L^*{}_{u2}$, and $L^*{}_{u3}$, with $y$ with that of the Kolmogorov scale, $L_k^*$, (black solid line) and the balance errors in the $\langle u^3 \rangle$-equation (black dash-dotted line). Red and blue lines are for the length scales at $n$ = 2 and 3, respectively. The scales $L^*{}_{un}$ and $L^*{}_{pn}$ are shown by the color dashed and solid lines.

The Kolmogorov scale is larger than $L^*{}_{inf}$ everywhere in the flow and larger than the other four scales at $y^+ > 5$. This makes $L^*{}_{inf}$ the safe choice for the reference scale in the areas of high mean velocity gradients. Whether the scales $L^*{}_{p2}$, $L^*{}_{p3}$, $L^*{}_{u2}$, and $L^*{}_{u3}$ being larger than the Kolmogorov scale at $y^+ < 5$ has an impact on the simulation results remains to be verified particularly in the light that the velocity moments are effectively suppressed in this flow area (Figs. 3-6).

The value of $L^*{}_{inf}$, 0.00255, is about twice larger than the threshold value of $L^*{}_{14}$ above which the balance errors become noticeable. From that, $L^*{}_{inf}/2$ is suggested as a criterion for the grid cell size to meet in the buffer zone and other flow areas with large velocity gradients.

The best reference lengthscale outside the areas of high mean velocity gradients has yet to be determined. On one hand, the scales $L^*{}_{u2}$, and $L^*{}_{u3}$ remain smaller than the Kolmogorov scale outside the flow buffer zone, but on the other hand, using the grid cell size less than the Kolmogorov lengthscale did not reduce the balance errors in [17] (see Fig. 11b). Separate studies focused on the analysis of uncertainty in DNS data in the flow areas of small mean velocity gradients will be beneficial.

## 4. Conclusion

In the paper, the model expressions for fifth-order velocity moments derived from the truncated Gram-Charlier series expansions model for a turbulent flow field probability density function were successfully validated in a planar incompressible flow in a strained channel with and without separation using DNS data for fifth and lower-order moments.

The imposed flow deformation was found to affect differently odd and even velocity moments, and the moments of different orders. In particular, the magnitude of the velocity moments which are the most and the least effected, are those in the streamwise direction, $\langle u^m \rangle$, and in the direction normal to the wall, $\langle v^m \rangle$, respectively. The odd moments are suppressed more than the even ones. The higher-order moments are suppressed stronger by the flow deformation than the lower-order ones. The most sensitive moment to the flow deformation is $\langle u^5 \rangle$, and the moment with the least affected maximum value is $\langle v^2 \rangle$, closely followed by $\langle v^4 \rangle$.

The imposed flow deformation was also found to be beneficial for suppressing a small discrepancy between the model and DNS profiles of the fifth-order velocity moments observed in the buffer zone of velocity moments containing velocity fluctuations in the streamwise direction. Accuracy of DNS profiles for odd velocity moments was identified as the cause of this phenomenon.

Further investigation revealed the DNS grid resolution in the buffer zone to be insufficient for particularly odd moments. A new length scale to improve the accuracy of higher-order statistics near walls and other areas of high velocity gradients is proposed.

## Acknowledgments

The authors would like to express their gratitude to Dr. E. Jeyapaul and Dr. G. N. Coleman (NASA Langley) for their effort into generating and documenting the DNS dataset unmatched in its completeness, and also to Dr. C. L. Rumsey (NASA Langley) for his effort in making this and other datasets available to the professional community. A progress in turbulence modeling, which includes understanding and mitigating uncertainties and errors in reference data requires a collective effort, and this dataset is an example of that.

## References

1. Kampé de Fériet. J. 1966 The Gram-Charlier approximation of the normal law and the statistical description of a homogeneous turbulent flow near statistical equilibrium. *David Taylor Model Basin Report* No. 2013, Naval Ship Research and Development Center, Washington D.C.

2. Appell, P., Kampé de Fériet. J. 1926 Fonctions Hypergéométriques et Hypersphériques. Polynomes d'Hermite. Gauthier-Villars et C^ie, Paris.

3. Barndorff-Nielsen, O., Pedersen, B. V. 1979 The bivariate Hermite polynomials up to order six. *Scand. J. Statist*. 6, 127-128.

4. Frenkiel, F. N, Klebanoff. P. S. 1973 Probability distributions and correlations in a turbulent boundary layer. *Phys. Fluids*. **16**(6), 725-737.

5. Durst. F., Jovanović, J., Johansson, T. G. 1992 On the statistical properties of truncated Gram–Charlier series expansions in turbulent wall-bounded flows. *Phys. Fluids A*, **4** (1), 118-127.

6. Millionshtchikov, M. D., 1941. On the theory of homogeneous isotropic turbulence. C. R. *Acad. Sci. SSSR*. **32**, 615-619.

7. Tsinober, A. 2001 *An informal introduction to turbulence*. Kluwer Academic Publishers.

8. Wilczek, M., Daitche, A., Friedrich, R. 2011 On the velocity distribution in homogeneous isotropic turbulence: correlations and deviations from Gaussianity. *J. Fluid Mech*. **676**, 191–217.

9. Jiménez, J. 1998 Turbulent velocity fluctuations need not be Gaussian. *J. Fluid Mech.* **376**, 139-147.

10. Tavoularis, S., Corrsin, S. 1981 Experiments in nearly homogenous turbulent shear flow with a uniform mean temperature gradient. Part 1. *J. Fluid Mech*. **104**, 311-347.

11. Dong, S., Lozano-Durán, A., Sekimoto, A., Jiménez, J. 2017, Coherent structures in statistically stationary homogeneous shear turbulence. *J. Fluid Mech*. **816**, 167-208.

12. Antonia, R. A, Atkinson, J. D. 1973 High-order moments of Reynolds shear stress fluctuations in a turbulent boundary layer. *J. Fluid Mech*. **58**(3), 581-593.

13. Nakagawa. H., Nezu, I. 1977 Prediction of the contributions to the Reynolds stress from bursting events in open-channel flows. *J. Fluid Mech*. **80**(1), 99-128.

14. Poroseva, S. V. 1996 *High-order turbulence closure in a fully-developed flow in a cylindrical pipe* (in Russian). Ph. D. Thesis, Novosibirsk State University, Novosibirsk, Russia.

15. Kurbatskii, A. F, Poroseva, S. V. 1997 A model for calculating the three components of the excess for the turbulent field of flow velocity in a round pipe rotating about its longitudinal axis. *High Temperature*. **35**(3), 432-440.

16. Poroseva, S. V, Murman, S. M. 2014 Velocity/pressure-gradient correlations in a FORANS approach to turbulence modeling. AIAA2014-2207, Proc. the AIAA Aviation and Aeronautics Forum and Exposition, Atlanta, GA, June 16-20.

17. Poroseva, S. V, Kaiser, B. E, Sillero, J. A., Murman, S. M. 2015 Validation of a closing procedure for fourth-order RANS turbulence models with DNS data in an incompressible zero-pressure-gradient turbulent boundary layer. *Int. J. Heat Fluid Flow*. **56**, 71-79.

18. Simens, M. P, Jiménez, J., Hoyas, S., Mizuno, Y. 2009 A high-resolution code for turbulent boundary layers. *J. Comp. Phys.* **228**(11), 4218-4231.

19. Borrell, G., Sillero, J. A., Jiménez, J. 2013 A code for direct numerical simulation of turbulent boundary layers at high Reynolds numbers in BG/P supercomputers. *Computers & Fluids*. **80**, 37-43.

20. Sillero, J. A., Jiménez, J., Moser, R. D. 2013 One-point statistics for turbulent wall-bounded flows at Reynolds numbers up to $\delta^+ = 2000$. *Phys. Fluids*. **25**(10), 1-15.

21. Jeyapaul, E., Coleman, G. N., Rumsey, C. L. 2015 Higher-order and length-scale statistics from DNS of a decelerated planar wall-bounded turbulent flow. *Int. J. Heat Fluid Flow*, **54**, 14-27.

22. Turbulence Modeling Resource, DNS: High-Order Moments in Unstrained and Strained Channel Flow, https://turbmodels.larc.nasa.gov/Other_DNS_Data/high-order-channel.html

23. Jeyapaul, E., Coleman, G. N., Rumsey, C. L. 2014 Assessment of higher-order RANS closures in a decelerated planar wall-bounded turbulent flow. AIAA 2014-2088, Proc. AIAA Aviation, Atlanta, GA June 16-20.

24. Coleman, G. N., Kim, J., Spalart, P. R. 2003 Direct numerical simulation of a decelerated wall-bounded turbulent shear flow. *J. Fluid Mech*. **495**, 1–18.

25. Yorke, C. P., Coleman, G. N. 2004 Assessment of common turbulence models for an idealised adverse pressure gradient flow. *European Journal of Mechanics B/Fluids*, 23, 319-337.

26. Sciberras, M. A., Coleman, G. N. 2007 Testing of Reynolds-stress-transport closures by comparison with DNS of an idealized adverse-pressure-gradient boundary layer. *European Journal of Mechanics B/Fluids*, **26**, 551-582.

27. Gence, J. N., Mathieu, J. 1979 On the application of successive plane strains to grid generated turbulence. J. Fluid Mech. **93**, 501-513.

28. Hunt, J. C. R., Carruthers, D. J. 1990. Rapid distortion theory and the “problems” of turbulence. J. Fluid Mech. **212**, 497–532

29. Reynolds, W. C., Kassinos, S. C. 1995. One-point modeling for rapidly deformed homogeneous turbulence. Proc. R. Soc. London Ser. A **451**, 87–104.

30. Kevlahan, N. K. R., Hunt, J. C. R. 1997. Nonlinear interactions in turbulence with strong irrotational straining. J. Fluid Mech. **337**, 333–364.

31. Cambon , C., Scott, J. F. 1999 Linear and non linear models of anisotropic turbulence. Ann. Rev. Fluid Mech. **31**, 1-53.

32. Gualtieri, P., Meneveau, C. 2010 Direct numerical simulations of turbulence subjected to a straining and destraining cycle. Phys. Fluids, **22**, 065104.

33. Zusi, C. J., Perot, J. B. 2013 Simulation and modeling of turbulence subjected to a period of uniform plane strain. Phys. Fluids, **25**, 110819.

34. Girimaji, S. S., Jeong, E., Poroseva, S. V. 2003 Pressure strain correlation in homogeneous anisotropic turbulence subjected to rapid strain dominated distortion. Phys. Fluids, **15**(10), 3209-3222.

35. Poroseva, S. V., Colmenares. F., J. D., Murman, S. M. 2016 On the accuracy of RANS simulations with DNS data,” *Phys. Fluids*. **28**(11). (doi: 10.1063/1.4966639)

36. Poroseva, S. V., Jeyapaul, E., Murman, S. M., Colmenares F., J. D. 2016 The effect of the DNS data averaging time on the accuracy of RANS-DNS simulations. AIAA2016-3940, Proc. AIAA Aviation, Washington DC June 13-17.

37. Kolmogorov, A. N. 1941 The local structure of turbulence in incompressible viscous fluid for very large Reynolds numbers. *Doklady Akad. Nauk SSSR*. **30**(4), 299-303. English translation by Levin V. 1991 “Turbulence and stochastic process: Kolmogorov's ideas 50 Years on. *Proc. R. Soc. Lond. A*, **434**(1890), 9-13.

38. Monin, A. S., Yaglom, A. M. 1963 On the laws of small-scale turbulent flow of liquids and gases. *Russ. Math. Surv*. **18**, 89-109.

39. Monin, A. S., Yaglom, A. M. 1979 *Statistical fluid mechanics: mechanics of turbulence. Vol. 1*. English Edition. (Ed. J. L. Lumley). The MIT Press.

40. Jiménez, J. 2018 Coherent structures in wall-bounded turbulence. J. Fluid Mech. **842**, 1-100.

## Appendix A

The two adjunct Hermite polynomials of two random variables:

$$H_{l,k}(s_1,s_2) = (-1)^{l+k} \exp\left[\frac{1}{2}\Phi(s_1,s_2)\right]\frac{\partial^{l+k}}{\partial s_1^l \partial s_2^k}\left\{\exp\left[-\frac{1}{2}\Phi(s_1,s_2)\right]\right\}, \quad \text{(A1)}$$

$$G_{l,k}(s_1,s_2) = (-1)^{l+k} \exp\left[\frac{1}{2}\Psi(\xi_1,\xi_2)\right]\frac{\partial^{l+k}}{\partial \xi_1^l \partial \xi_2^k}\left\{\exp\left[-\frac{1}{2}\Psi(\xi_1,\xi_2)\right]\right\}. \quad \text{(A2)}$$

are connected with two positive definite quadratic forms

$$\Phi(s_1,s_2) = \frac{s_1^2 - 2Rs_1s_2 + s_2^2}{1-R^2}, \quad \text{(A3)}$$

and

$$\Psi(\xi_1,\xi_2) = \xi_1^2 + 2R\xi_1\xi_2 + \xi_2^2, \quad \text{(A4)}$$

with

$$\xi_1 = \frac{s_1 - Rs_2}{1-R^2}, \xi_2 = \frac{s_2 - Rs_1}{1-R^2} \quad \text{(A5)}$$

(and thus, $\Phi(s_1,s_2) = \Psi(\xi_1,\xi_2)$) [1]. In a turbulent flow with $s_1 = u_1/\sigma_1$ and $s_2 = u_2/\sigma_2$, the parameter $R$ corresponds to the standardized statistical velocity moment $R^{m,n}(s_1,s_2) = \langle s_1^m s_2{}^n \rangle = \langle \left(\frac{u_1}{\sigma_1}\right)^m \left(\frac{u_2}{\sigma_2}\right)^n \rangle = \frac{\langle u_1^m u_2{}^n \rangle}{\sigma_1^m \sigma_2^n}$ with $m = n = 1$: $R = R^{1,1}$, where $\langle u_1^m u_2^n \rangle$ are statistical velocity moments of the $(m+n)$ order, and $u_1$ and $u_2$ can be any component of the turbulent velocity vector $\vec{u} = (u,v,w)$ (in the Cartesian coordinates).

From (A2), (A4), and (A5), one can obtain explicit expressions for $G_{l,k}$ in terms of random variables $\xi_1$ and $\xi_2$ and the standardized moment $R$, and then, substitute the ξ-variables with the $s$-variables. The procedure is time-consuming. In [1], the polynomials for $l + k \leq 4$ were obtained. In [3], the expressions for $G_{l,k}$ were derived in a more general form (not limited to turbulent flows) for the polynomials up to $l + k = 6$. Since corrections were found to be necessary for $G_{1,1}$ and $G_{2,2}$ in [1] and additional steps have to be performed on the expressions in [3] to make them applicable to the framework developed in [1] (also a time-consuming process), here the corrected $G$ -polynomials to the fourth order are provided. First, expressions at $l + k \leq 2$ are given in both sets of variables to clarify how they are relevant to each other:

$$G_{0,0} = 1,$$
$$G_{1,0} = \xi_1 + R\xi_2 = s_1, \qquad G_{0,1} = \xi_2 + R\xi_1 = s_2,$$
$$G_{2,0} = \xi_1^2 + 2R\xi_1\xi_2 + R^2\xi_2^2 - 1 = s_1^2 - 1, \; G_{0,2} = s_2^2 - 1, \tag{A6}$$
$$G_{1,1} = R\xi_1^2 + (R^2 + 1)\xi_1\xi_2 + R\xi_2^2 - R = s_1 s_2 - R.$$

Expressions (A6) written in the $\xi$-variables correspond to those in [3] obtained with the substitutions $H \to G$ and $(s_1, s_2) \to (\xi_1, \xi_2)$ and with the positive definite matrix elements $\delta_{11} = \delta_{22} = 1$ and $\delta_{12} = R$ derived in [1].

The expressions for $2 < l + k \leq 4$ are given only in the $s$-variables as the corresponding expressions in the $\xi$-variables become too large and will not be used further:

$$G_{3,0} = s_1^3 - 3s_1, \qquad G_{0,3} = s_2^3 - 3s_2,$$
$$G_{2,1} = s_1^2 s_2 - s_2, \qquad G_{1,2} = s_1 s_2^2 - s_1,$$
$$G_{4,0} = s_1^4 - 6s_1^2 + 3, \; G_{3,1} = s_1^3 s_2 - 3s_1 s_2 + 3Rs_1^2 - 3R, \tag{A7}$$

$$G_{2,2} = s_1^2 s_2^2 - s_1^2 - s_2^2 - 4Rs_1 s_2 + 2R^2 + 1,$$

$$G_{1,3} = s_1 s_2^3 - 3s_1 s_2 + 3Rs_2^2 - 3R, \qquad G_{0,4} = s_2^4 - 6s_2^2 + 3.$$

Using expressions (A6)-(A7) and alike for higher orders in the definition of the coefficients $A_{l,k}$ in (1), one can obtain expressions for these coefficients in terms of the standardized velocity moments. Here, the coefficients obtained using expressions (A6) and (A7) are provided:

$$A_{0,0} = 1,$$

$$A_{1,0} = 0, \qquad A_{0,1} = 0,$$

$$A_{2,0} = 0, \qquad A_{1.1} = 0, \qquad A_{0,2} = 0,$$

$$A_{3,0} = \frac{R^{3,0}}{6}, \qquad A_{2,1} = \frac{R^{2,1}}{2}, \qquad A_{1,2} = \frac{R^{1,2}}{2}, \qquad A_{0,3} = \frac{R^{0,3}}{6}, \tag{A8}$$

$$A_{4,0} = \frac{R^{4,0} - 3}{24}, \qquad A_{3,1} = \frac{R^{3,1} - 3R}{6}, \qquad A_{2,2} = \frac{R^{2,2} - R^{2,0} - R^{0.2} - 2R^2 + 1}{4},$$

$$A_{1,3} = \frac{R^{1,3} - 3R}{6}, \qquad A_{0,4} = \frac{R^{0.4} - 3}{24}.$$

In turbulence, $R^{2,0} = R^{0,2} = 1$ by definition, which yields:

$$A_{2,2} = \frac{R^{2,2} - 2R^2 - 1}{4}.$$

## Appendix B

This section provides transport equations for velocity moments that appear in a fourth-order statistical closure in the channel flow considered in the current study prior any modeling:

$$\frac{\partial\langle v^2\rangle}{\partial t}+V\frac{\partial\langle v^2\rangle}{\partial y}=2\langle v^2\rangle B_{11}-\frac{\partial\langle v^3\rangle}{\partial y}+\nu\frac{\partial^2\langle v^2\rangle}{\partial y\partial y}+\Pi_{yy}-\varepsilon_{yy}, \tag{B1}$$

$$\frac{\partial\langle uv\rangle}{\partial t}+V\frac{\partial\langle uv\rangle}{\partial y}=-\langle v^2\rangle\frac{\partial U}{\partial y}-\frac{\partial\langle uv^2\rangle}{\partial y}+\nu\frac{\partial^2\langle uv\rangle}{\partial y\partial y}+\Pi_{xy}-\varepsilon_{xy}, \tag{B2}$$

$$\frac{\partial\langle v^3\rangle}{\partial t}+V\frac{\partial\langle v^3\rangle}{\partial y}=3\langle v^3\rangle B_{11}-\frac{\partial\langle v^4\rangle}{\partial y}+\nu\frac{\partial^2\langle v^3\rangle}{\partial y\partial y}+3\langle v^2\rangle\frac{\partial\langle v^2\rangle}{\partial y}+\Pi_{yyy}-\varepsilon_{yyy}, \tag{B3}$$

$$\frac{\partial\langle uv^2\rangle}{\partial t}+V\frac{\partial\langle uv^2\rangle}{\partial y}=\langle uv^2\rangle B_{11}-\langle v^3\rangle\frac{\partial U}{\partial y}-\frac{\partial\langle uv^3\rangle}{\partial y}+\nu\frac{\partial^2\langle uv^2\rangle}{\partial y\partial y}+\langle v^2\rangle\frac{\partial\langle uv\rangle}{\partial y}+2\langle uv\rangle\frac{\partial\langle v^2\rangle}{\partial y}+\Pi_{xyy}-\varepsilon_{xyy}, \tag{B4}$$

$$\frac{\partial\langle v^4\rangle}{\partial t}+V\frac{\partial\langle v^4\rangle}{\partial y}=4\langle v^4\rangle B_{11}-\frac{\partial\langle v^5\rangle}{\partial y}+\nu\frac{\partial^2\langle v^4\rangle}{\partial y\partial y}+4\langle v^3\rangle\frac{\partial\langle v^2\rangle}{\partial y}+\Pi_{yyyy}-\varepsilon_{yyyy}, \tag{B5}$$

$$\begin{aligned}\frac{\partial\langle uv^3\rangle}{\partial t}+V\frac{\partial\langle uv^3\rangle}{\partial y}&=2\langle uv^3\rangle B_{11}-\langle v^4\rangle\frac{\partial U}{\partial y}-\frac{\partial\langle uv^4\rangle}{\partial y}+\nu\frac{\partial^2\langle uv^3\rangle}{\partial y\partial y}+\langle v^3\rangle\frac{\partial<uv\rangle}{\partial y}+\\&+3\langle uv^2\rangle\frac{\partial\langle v^2\rangle}{\partial y}+\Pi_{xyyy}-\varepsilon_{xyyy}.\end{aligned} \tag{B6}$$

These equations were obtained from the general transport equation for the velocity moment $<u^n v^m w^k>$ of arbitrary order in an incompressible planar turbulent flow [35] taking into account the flow homogeneity in the *x*- and *z*-directions preserved under the imposed flow deformation. The governing equations can also be obtained from those provided in [24].

To close this set of six equations, one has to model two fifth-order velocity moments: $\langle uv^4 \rangle$ and $\langle v^5 \rangle$ and some of the components of the tensors $\mathbf{\Pi}$ and $\boldsymbol{\varepsilon}$ (six of each) describing interaction of the turbulent velocity and pressure fields (velocity/pressure-gradient correlations) and the turbulent kinetic energy dissipation, respectively.

Models (5) for $\langle uv^4 \rangle$ and $\langle v^5 \rangle$: $\langle uv^4 \rangle = 6\langle v^2 \rangle \cdot \langle uv^2 \rangle + 4\langle v^3 \rangle \cdot \langle uv \rangle$ and $\langle v^5 \rangle = 10\langle v^2 \rangle \cdot \langle v^3 \rangle$, **do not bring additional unknown parameters** into the set of equations (B1)-(B6), which is another benefit of this modeling approach. However, models for unknown components of the tensors $\mathbf{\Pi}$ and $\boldsymbol{\varepsilon}$ (not considered here) usually rely on variables that are not originally present in (B1)-(B6). As a result, the original set of equations may expand to include equations for new unknown parameters. For this reason, Appendix C provides transport equations for all velocity moments of the fourth order that may be found in this flow. Equations for velocity moments $\langle u^2 \rangle$ and $\langle u^3 \rangle$ are provided below:

$$\frac{\partial \langle u^2 \rangle}{\partial t} + V\frac{\partial \langle u^2 \rangle}{\partial y} = -2\langle u^2 \rangle B_{11} - 2\langle uv \rangle \frac{\partial U}{\partial y} - \frac{\partial \langle u^2 v \rangle}{\partial y} + \nu \frac{\partial^2 \langle u^2 \rangle}{\partial y \partial y} + \Pi_{xx} - \varepsilon_{xx}, \qquad \text{(B7)}$$

$$\frac{\partial \langle u^3 \rangle}{\partial t} + V\frac{\partial \langle u^3 \rangle}{\partial y} = -3\langle u^3 \rangle B_{11} - 3\langle u^2 v \rangle \frac{\partial U}{\partial y} - \frac{\partial \langle u^3 v \rangle}{\partial y} + 3\langle u^2 \rangle \frac{\partial \langle uv \rangle}{\partial y} + \nu \frac{\partial^2 \langle u^3 \rangle}{\partial y \partial y} + \Pi_{xxx} - \varepsilon_{xxx}. \qquad \text{(B8)}$$

Expressions for unknown components of the tensors $\mathbf{\Pi}$ and $\boldsymbol{\varepsilon}$ in (B1)-(B6) are the following:

$$\Pi_{xx} = -\frac{2}{\rho}\langle x\frac{\partial p}{\partial x}\rangle,\ \Pi_{yy} = -\frac{2}{\rho}\langle v\frac{\partial p}{\partial y}\rangle,\ \Pi_{xy} = -\frac{1}{\rho}\left[\langle v\frac{\partial p}{\partial x}\rangle + \langle u\frac{\partial p}{\partial y}\rangle\right],$$

$$\Pi_{xxx} = -\frac{3}{\rho}\langle u^2\frac{\partial p}{\partial x}\rangle,\ \ \Pi_{xyy} = -\frac{1}{\rho}\left[\langle v^2\frac{\partial p}{\partial x}\rangle + 2\langle uv\frac{\partial p}{\partial y}\rangle\right],\ \ \Pi_{yyy} = -\frac{3}{\rho}\langle v^2\frac{\partial p}{\partial y}\rangle,$$

$$\Pi_{xyyy} = -\frac{1}{\rho}\left[\langle v^3\frac{\partial p}{\partial x}\rangle + 3\langle uv^2\frac{\partial p}{\partial y}\rangle\right],\quad \Pi_{yyyy} = -\frac{4}{\rho}\langle v^3\frac{\partial p}{\partial y}\rangle,$$

$$\varepsilon_{xx} = 2\nu\langle\left[\left(\frac{\partial u}{\partial x}\right)^2 + \left(\frac{\partial u}{\partial y}\right)^2 + \left(\frac{\partial u}{\partial z}\right)^2\right]\rangle,\ \ \varepsilon_{yy} = 2\nu\langle\left[\left(\frac{\partial v}{\partial x}\right)^2 + \left(\frac{\partial v}{\partial y}\right)^2 + \left(\frac{\partial v}{\partial z}\right)^2\right]\rangle,$$

$$\varepsilon_{xy} = 2\nu\langle\left[\frac{\partial u}{\partial x}\frac{\partial v}{\partial x} + \frac{\partial u}{\partial y}\frac{\partial v}{\partial y} + \frac{\partial u}{\partial z}\frac{\partial v}{\partial z}\right]\rangle,\ \varepsilon_{xxx} = 6\nu\langle u\left[\left(\frac{\partial u}{\partial x}\right)^2 + \left(\frac{\partial u}{\partial y}\right)^2 + \left(\frac{\partial u}{\partial z}\right)^2\right]\rangle,$$

$$\varepsilon_{xyy} = 4\nu\langle v\left[\frac{\partial u}{\partial x}\frac{\partial v}{\partial x} + \frac{\partial u}{\partial y}\frac{\partial v}{\partial y} + \frac{\partial u}{\partial z}\frac{\partial v}{\partial z}\right]\rangle + 2\nu\langle u\left[\left(\frac{\partial v}{\partial x}\right)^2 + \left(\frac{\partial v}{\partial y}\right)^2 + \left(\frac{\partial v}{\partial z}\right)^2\right]\rangle,$$

$$\varepsilon_{yyy} = 6\nu\langle v\left[\left(\frac{\partial v}{\partial x}\right)^2 + \left(\frac{\partial v}{\partial y}\right)^2 + \left(\frac{\partial v}{\partial z}\right)^2\right]\rangle,\ \varepsilon_{yyyy} = 12\nu\langle v^2\left[\left(\frac{\partial v}{\partial x}\right)^2 + \left(\frac{\partial v}{\partial y}\right)^2 + \left(\frac{\partial v}{\partial z}\right)^2\right]\rangle,$$

$$\varepsilon_{xyyy} = 6\nu\left(\langle v^2\left[\frac{\partial u}{\partial x}\frac{\partial v}{\partial x} + \frac{\partial u}{\partial y}\frac{\partial v}{\partial y} + \frac{\partial u}{\partial z}\frac{\partial v}{\partial z}\right]\rangle + \langle uv\left[\left(\frac{\partial v}{\partial x}\right)^2 + \left(\frac{\partial v}{\partial y}\right)^2 + \left(\frac{\partial v}{\partial z}\right)^2\right]\rangle\right).$$

Here, $p$ is pressure fluctuation, $\rho$ is the fluid density.

## Appendix C

The set of RANS equations for the fourth-order velocity moments in the strained channel flow is the following:

$$\frac{\bar{D}\langle u^4\rangle}{Dt} = -4\langle u^4\rangle B_{11} - 4\langle u^3 v\rangle\frac{\partial U}{\partial y} - \frac{\partial\langle u^4 v\rangle}{\partial y} + 4\langle u^3\rangle\frac{\partial\langle uv\rangle}{\partial y} + \nu\frac{\partial^2\langle u^4\rangle}{\partial y\partial y} + \Pi_{xxxx} - \varepsilon_{xxxx}, \quad \text{(C1)}$$

$$\frac{\bar{D}\langle u^3 v\rangle}{Dt} = -\frac{\partial\langle u^3 v^2\rangle}{\partial y} - 2\langle u^3 v\rangle B_{11} - 3\langle u^2 v^2\rangle\frac{\partial U}{\partial y} + 3\langle u^2 v\rangle\frac{\partial\langle uv\rangle}{\partial y} + \nu\frac{\partial^2\langle u^3 v\rangle}{\partial y\partial y} + \Pi_{xxxy} - \varepsilon_{xxxy}, \quad \text{(C2)}$$

$$\frac{\bar{D}\langle u^2 v^2\rangle}{Dt} = -\frac{\partial\langle u^2 v^3\rangle}{\partial y} - 2\langle uv^3\rangle\frac{\partial U}{\partial y} + 2\langle uv^2\rangle\frac{\partial\langle uv\rangle}{\partial y} + 2\langle u^2 v\rangle\frac{\partial\langle v^2\rangle}{\partial y} + \nu\frac{\partial^2\langle u^2 v^2\rangle}{\partial y\partial y} + \Pi_{xxyy} - \varepsilon_{xxyy}, \quad \text{(C3)}$$

$$\frac{\bar{D}\langle uv^3\rangle}{Dt} = -\frac{\partial\langle uv^4\rangle}{\partial y} + 2\langle uv^3\rangle B_{11} - \langle v^4\rangle\frac{\partial U}{\partial y} + \langle v^3\rangle\frac{\partial\langle uv\rangle}{\partial y} + 3\langle uv^2\rangle\frac{\partial\langle v^2\rangle}{\partial y} + \nu\frac{\partial^2\langle uv^3\rangle}{\partial y\partial y} + \Pi_{xyyy} - \varepsilon_{xyyy}, \quad \text{(C4)}$$

$$\frac{\bar{D}\langle v^4\rangle}{Dt} = -\frac{\partial\langle v^5\rangle}{\partial y} + 4\langle v^4\rangle B_{11} + 4\langle v^3\rangle\frac{\partial\langle v^2\rangle}{\partial y} + \nu\frac{\partial^2\langle v^4\rangle}{\partial y\partial y} + \Pi_{yyyy} - \varepsilon_{yyyy}. \quad \text{(C5)}$$

Here, $\frac{\bar{D}}{Dt} = \frac{\partial}{\partial t} + V\frac{\partial}{\partial y}$. Unknown components of the fourth-rank tensors $\mathbf{\Pi}$ and $\boldsymbol{\varepsilon}$ in equations (C1)-(C5) are expressed as:

$$\Pi_{xxxx} = -\frac{4}{\rho}\langle u^3\frac{\partial p}{\partial x}\rangle, \quad \Pi_{xxxy} = -\frac{1}{\rho}\left[3\langle u^2 v\frac{\partial p}{\partial x}\rangle + \langle u^3\frac{\partial p}{\partial y}\rangle\right], \quad \Pi_{xxyy} = -\frac{2}{\rho}\left[\langle uv^2\frac{\partial p}{\partial x}\rangle + \langle u^2 v\frac{\partial p}{\partial y}\rangle\right],$$

$$\varepsilon_{xxxx} = 12\nu\langle u^2\left[\left(\frac{\partial u}{\partial x}\right)^2 + \left(\frac{\partial u}{\partial y}\right)^2 + \left(\frac{\partial u}{\partial z}\right)^2\right]\rangle, \qquad \varepsilon_{yyyy} = 12\nu\langle v^2\left[\left(\frac{\partial v}{\partial x}\right)^2 + \left(\frac{\partial v}{\partial y}\right)^2 + \left(\frac{\partial v}{\partial z}\right)^2\right]\rangle,$$

$$\varepsilon_{xxxy} = 6\nu\left(\langle u^2\left[\frac{\partial u}{\partial x}\frac{\partial v}{\partial x} + \frac{\partial u}{\partial y}\frac{\partial v}{\partial y} + \frac{\partial u}{\partial z}\frac{\partial v}{\partial z}\right]\rangle + \langle uv\left[\left(\frac{\partial u}{\partial x}\right)^2 + \left(\frac{\partial u}{\partial y}\right)^2 + \left(\frac{\partial u}{\partial z}\right)^2\right]\rangle\right),$$

$$\varepsilon_{xxyy} = 2\nu\left(\langle v^2\left[\left(\frac{\partial u}{\partial x}\right)^2 + \left(\frac{\partial u}{\partial y}\right)^2 + \left(\frac{\partial u}{\partial z}\right)^2\right]\rangle + \langle u^2\left[\left(\frac{\partial v}{\partial x}\right)^2 + \left(\frac{\partial v}{\partial y}\right)^2 + \left(\frac{\partial v}{\partial z}\right)^2\right]\rangle\right)$$
$$+8\nu\langle uv\left[\frac{\partial u}{\partial x}\frac{\partial v}{\partial x} + \frac{\partial u}{\partial y}\frac{\partial v}{\partial y} + \frac{\partial u}{\partial z}\frac{\partial v}{\partial z}\right]\rangle,$$

Other terms can be found in Appendix B.

## FIGURES

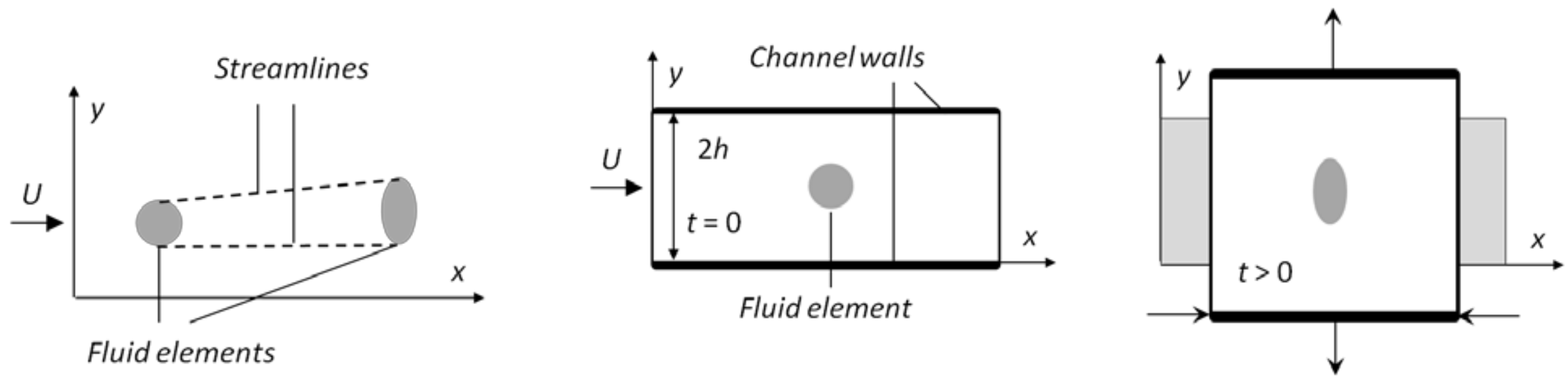

**a)** **b)**

FIG. 1. Side view of a planar turbulent APG boundary layer: a) spatially developing flow, b) time-evolving flow approximation in a channel: initial at $t = 0$ and deformed at $t > 0$.

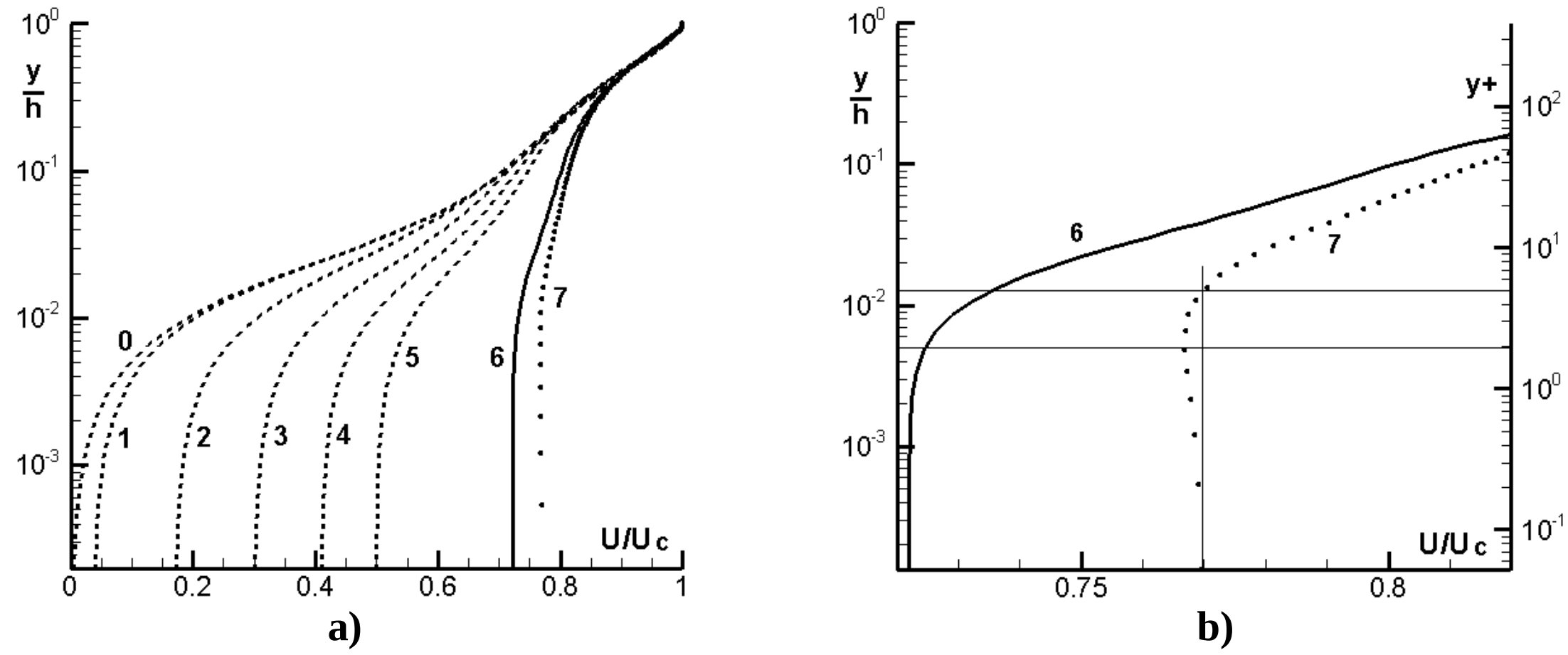

FIG. 2. The mean velocity profiles in a time-evolving strained planar turbulent channel flow at different values of $t'$: $0, 0.02, 0.1, 0.191, 0.285, 0.365, 0.675,$ and $0.772$ (shown from left to right). Labels 0, 3, 5, 6, and 7 correspond to $t_i'$, where $i = 0,3,5,6,$ and 7. Line patterns: dotted – $t_7'$, solid – $t_6'$, dashed – pre-separated flow at earlier times indicated by the labels. Horizontal thin lines in Fig. 2b show approximate locations of the separation region and of the velocity minimum within the flow.

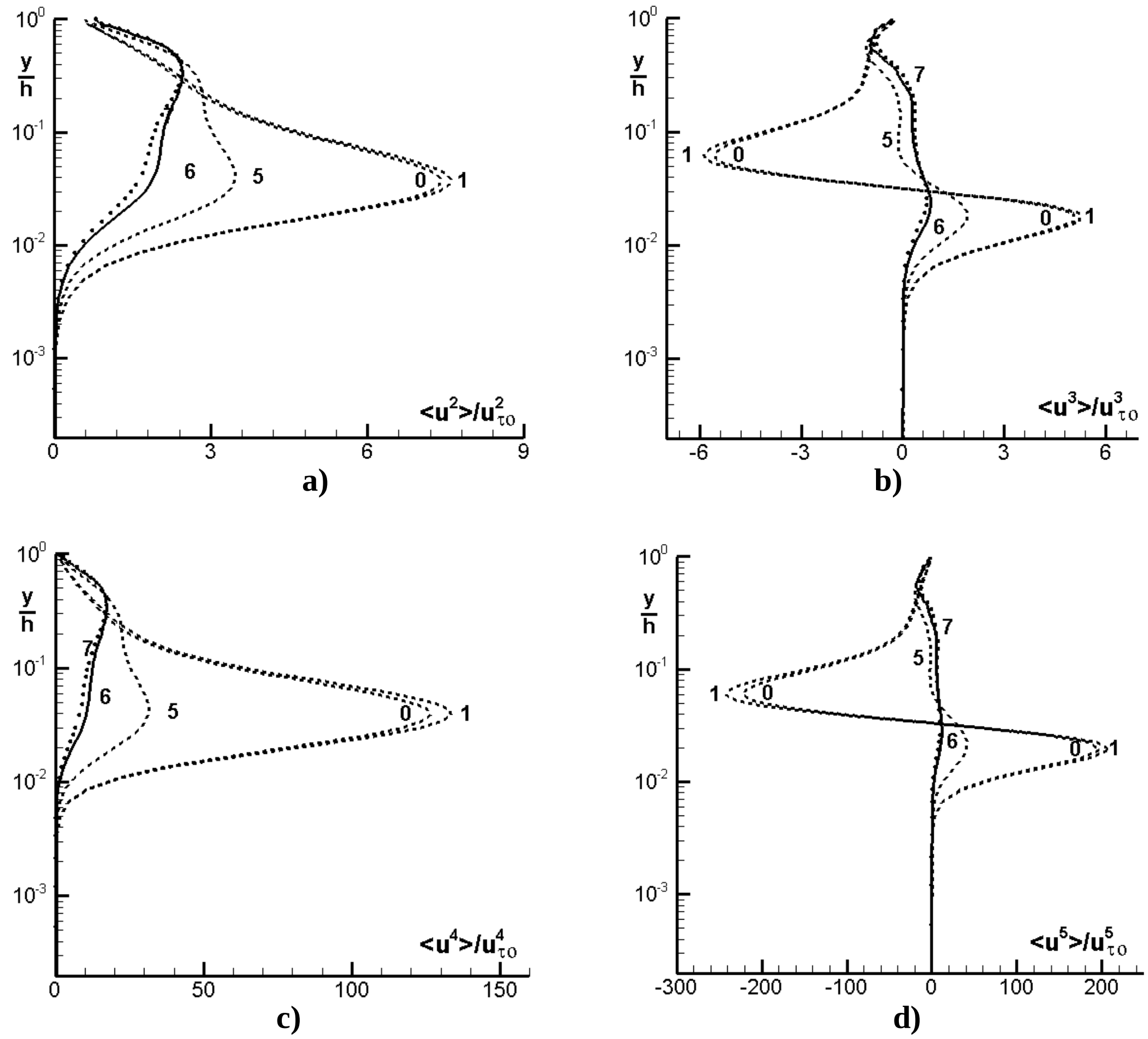

FIG. 3. Variation of $\langle u^m \rangle, m = 2, \ldots, 5$ with time in a time-evolving strained planar turbulent channel flow. Notations are the same as in Fig. 2.

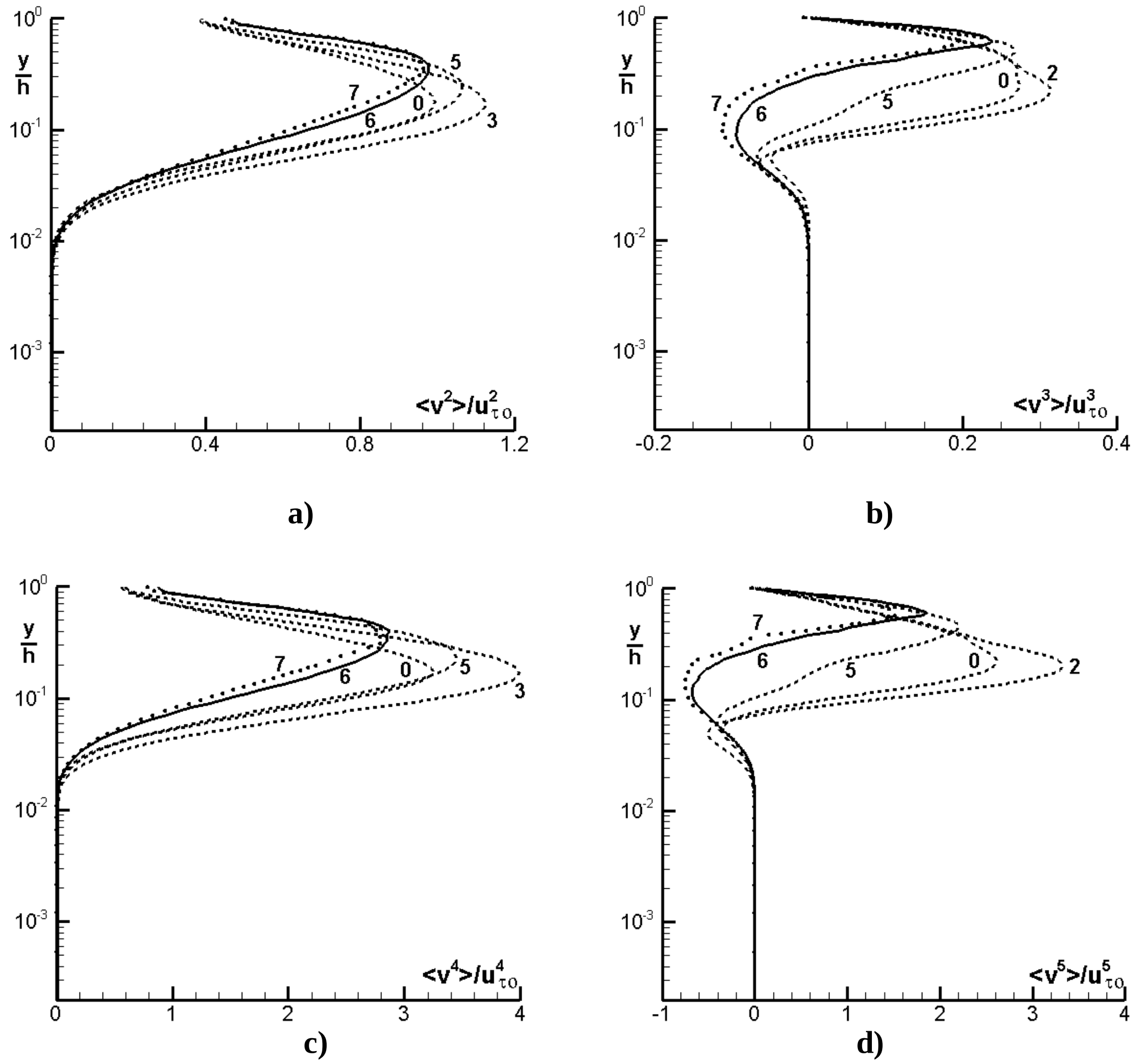

**a)** **b)** **c)** **d)**

FIG. 4. Variation of $\langle v^m \rangle, m = 2, \ldots, 5$ with time in a time-evolving strained planar turbulent channel flow. Notations are the same as in Fig. 2.

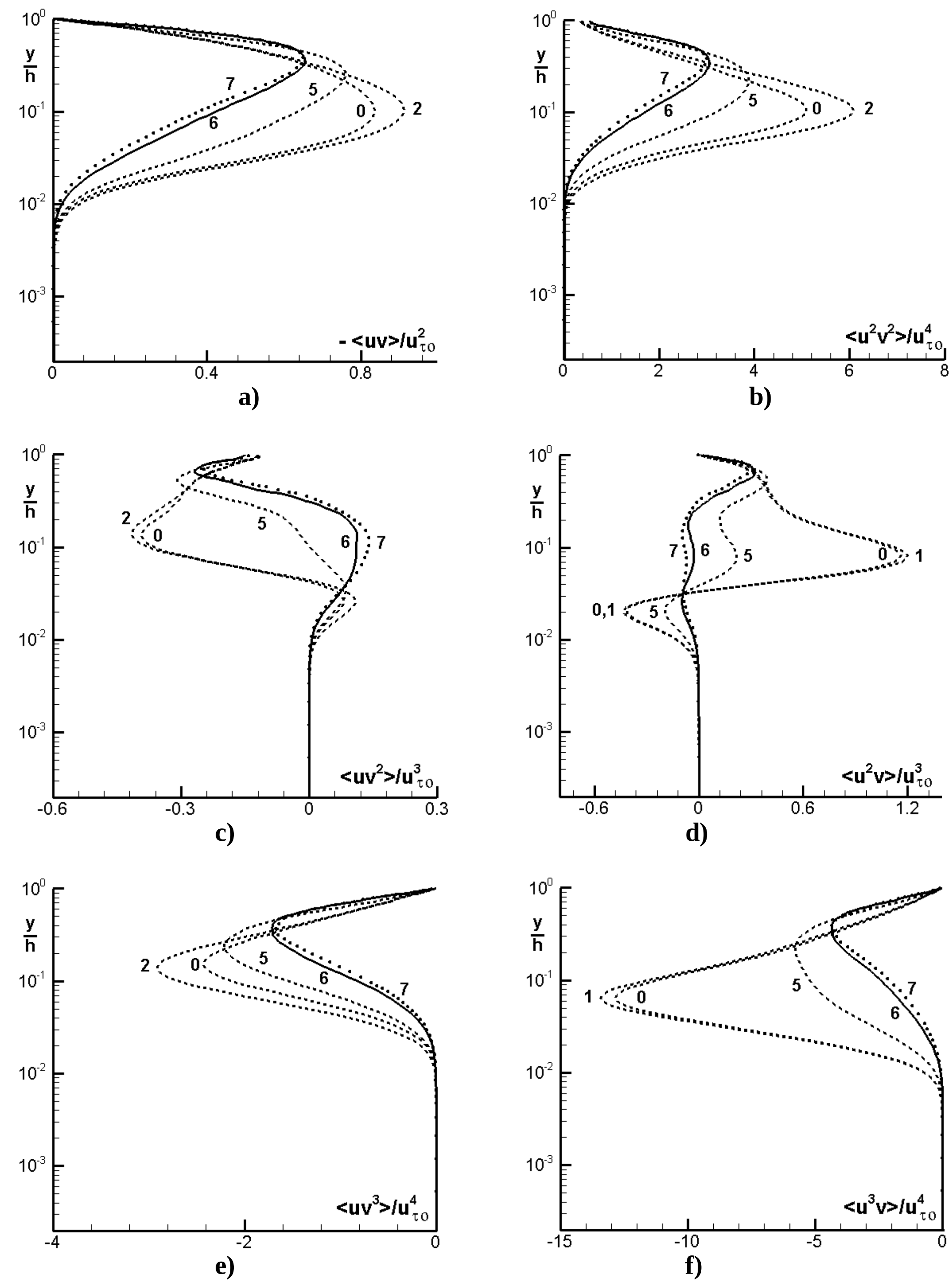

FIG. 5. Variation of $\langle u^m v^n \rangle (\, 2 \leq m+n \leq 4,\ m,n \geq 1)$ with time in a time-evolving strained planar turbulent channel flow. Notations are the same as in Fig. 2.

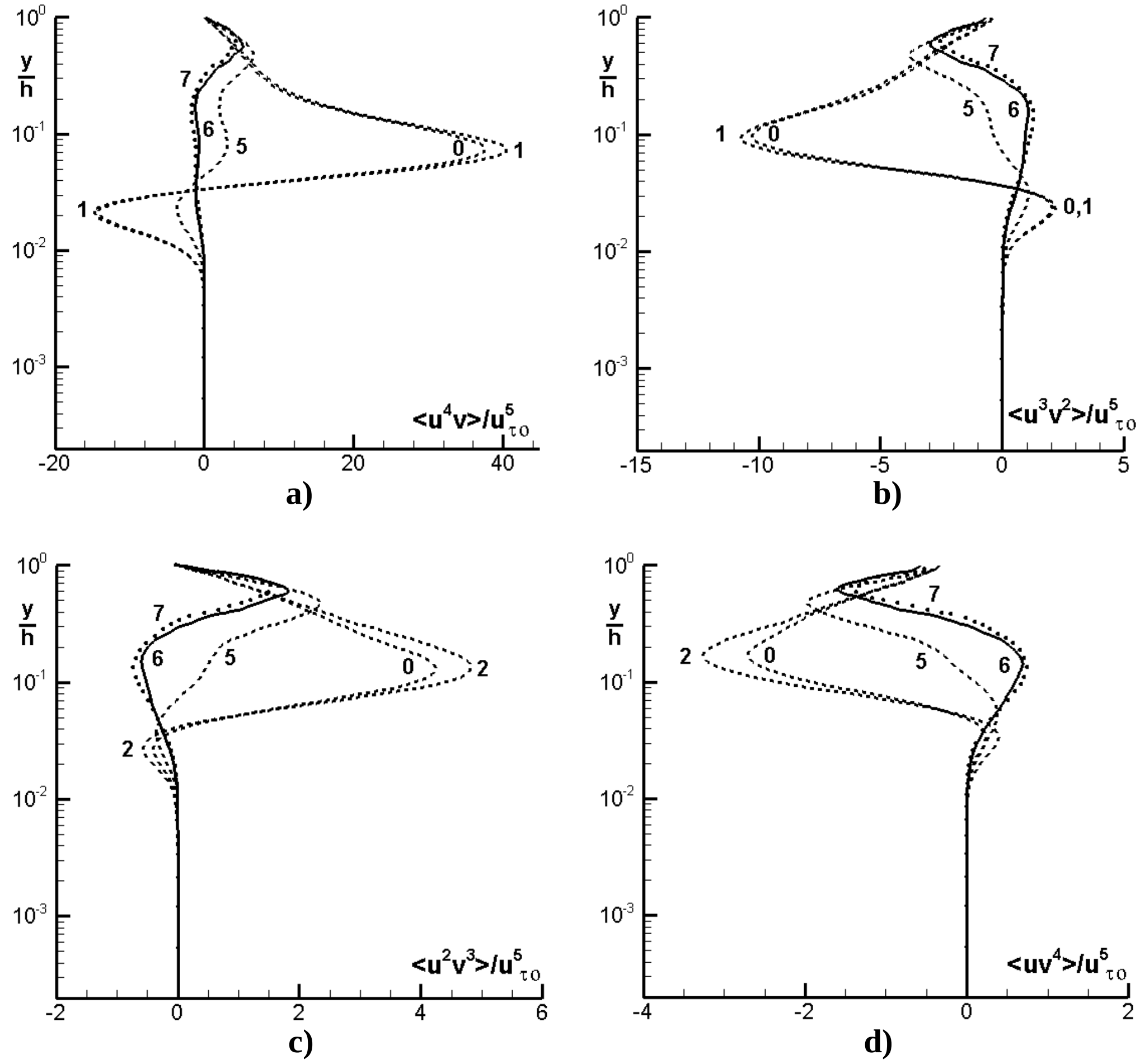

FIG. 6. Variation of $\langle u^m v^n \rangle (m = 1, \ldots, 4;\ n = 5 - m)$ with time in a time-evolving strained planar turbulent channel flow. Notations are the same as in Fig. 2.

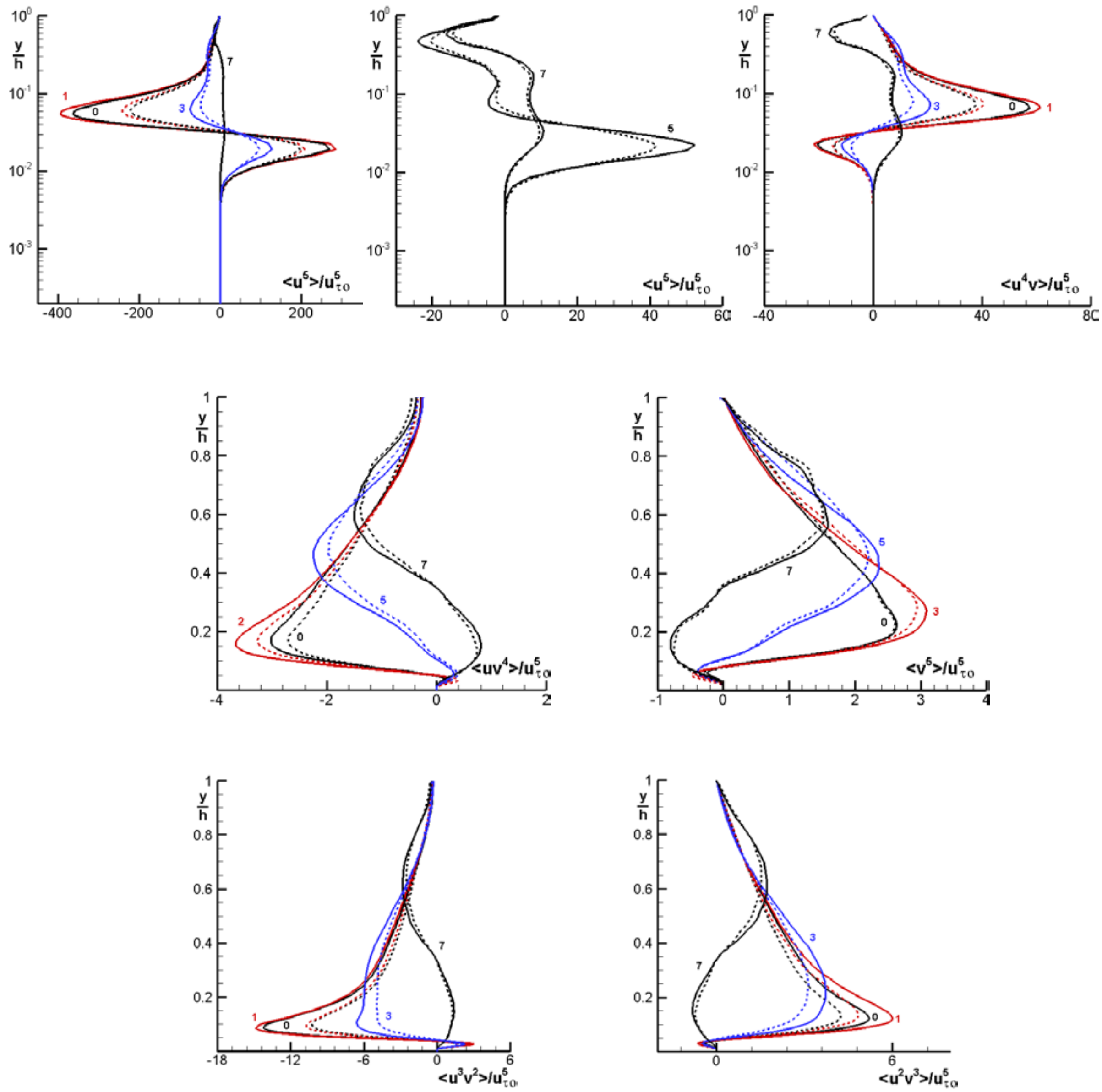

FIG. 7. The fifth-order moments: solid lines – profiles obtained from expressions (6) using DNS data [21,22] for the second- and third-order moments, dashed lines – DNS data for the fifth moments [21,22]. Labels are the same as in Fig.2. Color scheme: black – profiles in the unstrained channel and in the channel flow with separation, red – the maximum profile for the given moment, blue – profiles after reaching the maximum and before the flow separation.

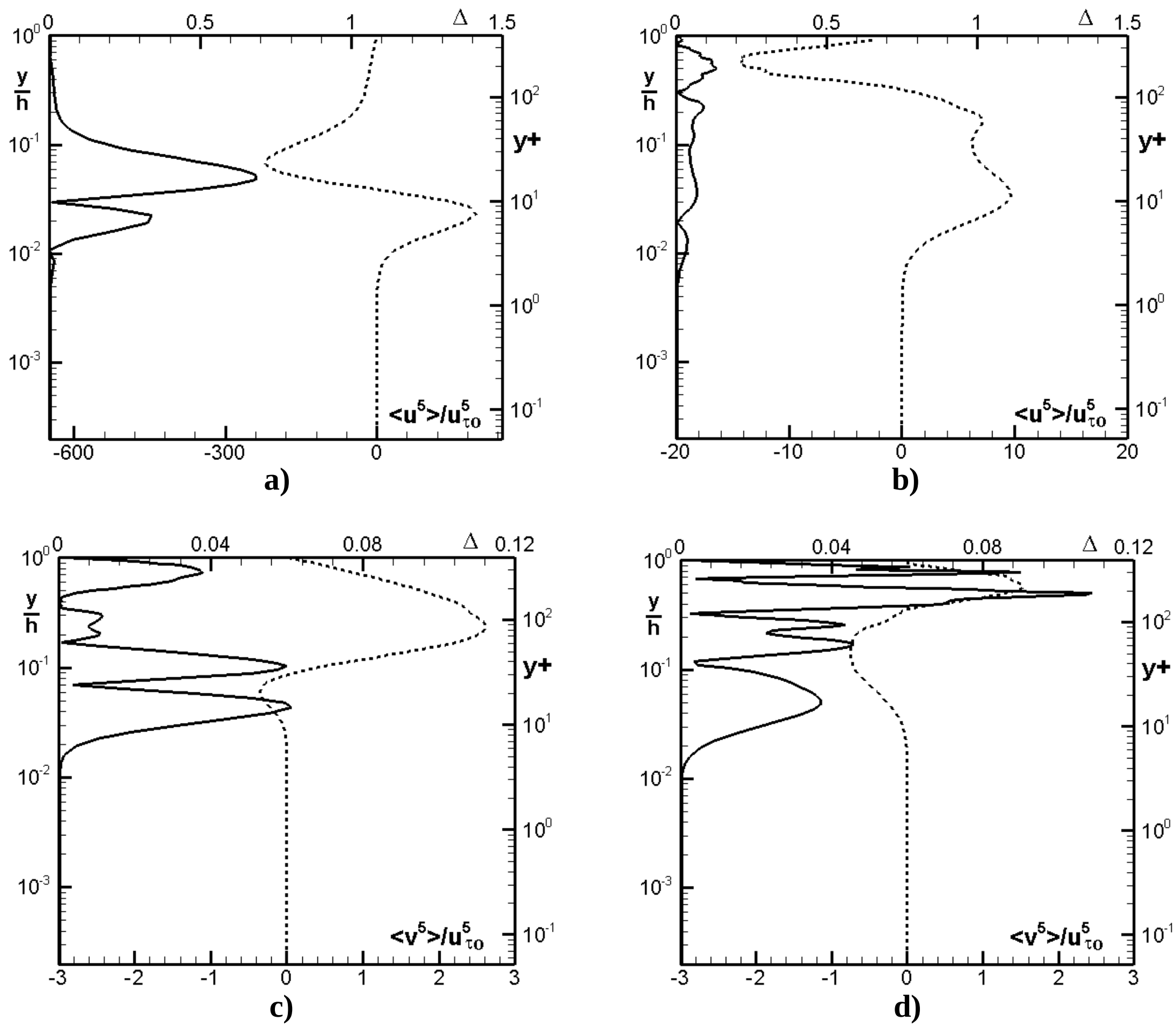

FIG. 8. The ratio Δ (Eq. (9)) for $\langle u^5 \rangle$ (a,b) and $\langle v^5 \rangle$ (c,d) at $t_0{}'$ (a,c) and $t_7{}'$ (b,d). Notation: solid lines – the ratios Δ; dashed lines – velocity moments.

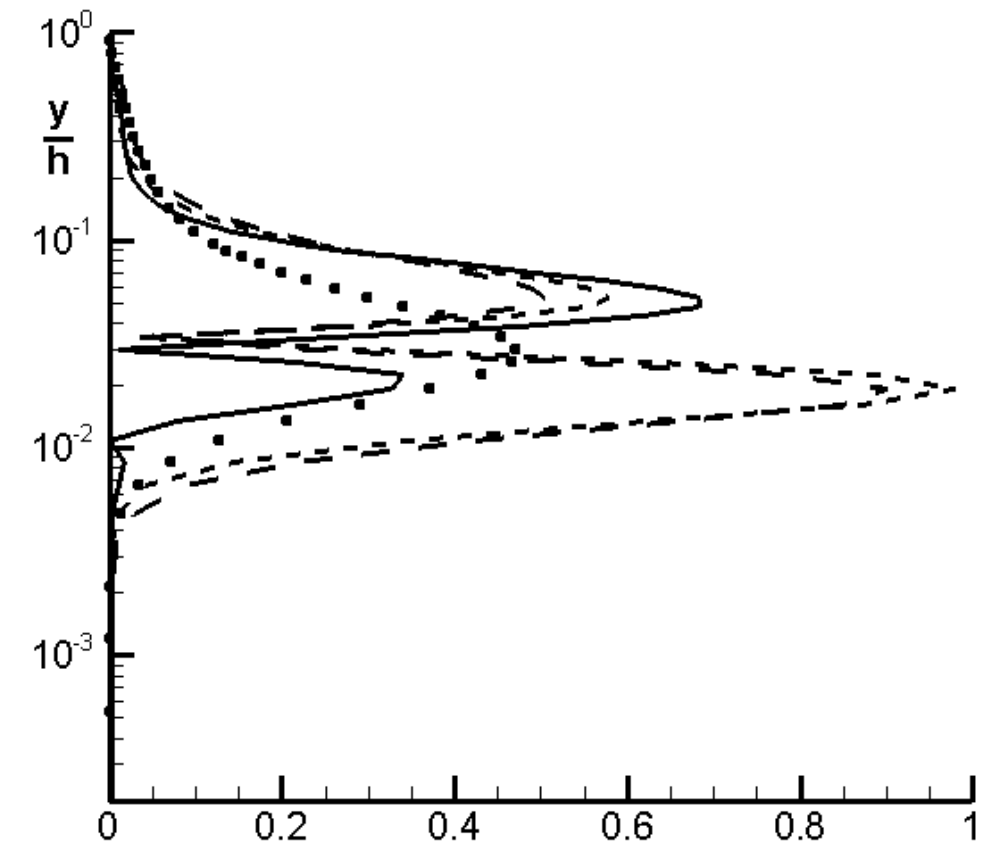

FIG. 9. Variation of the ratio Δ (Eq. (9)) for $\langle u^5 \rangle$ and the turbulence production by $\partial U/\partial y$ (the production terms are not to scale). Notations: solid line – $\langle u^5 \rangle$, dashed lines – production terms in the transport equations for $\langle u^5 \rangle$ and $\langle u^3 \rangle$ (absolute values), dotted line – production term in the transport equations for $\langle u^2 \rangle$.

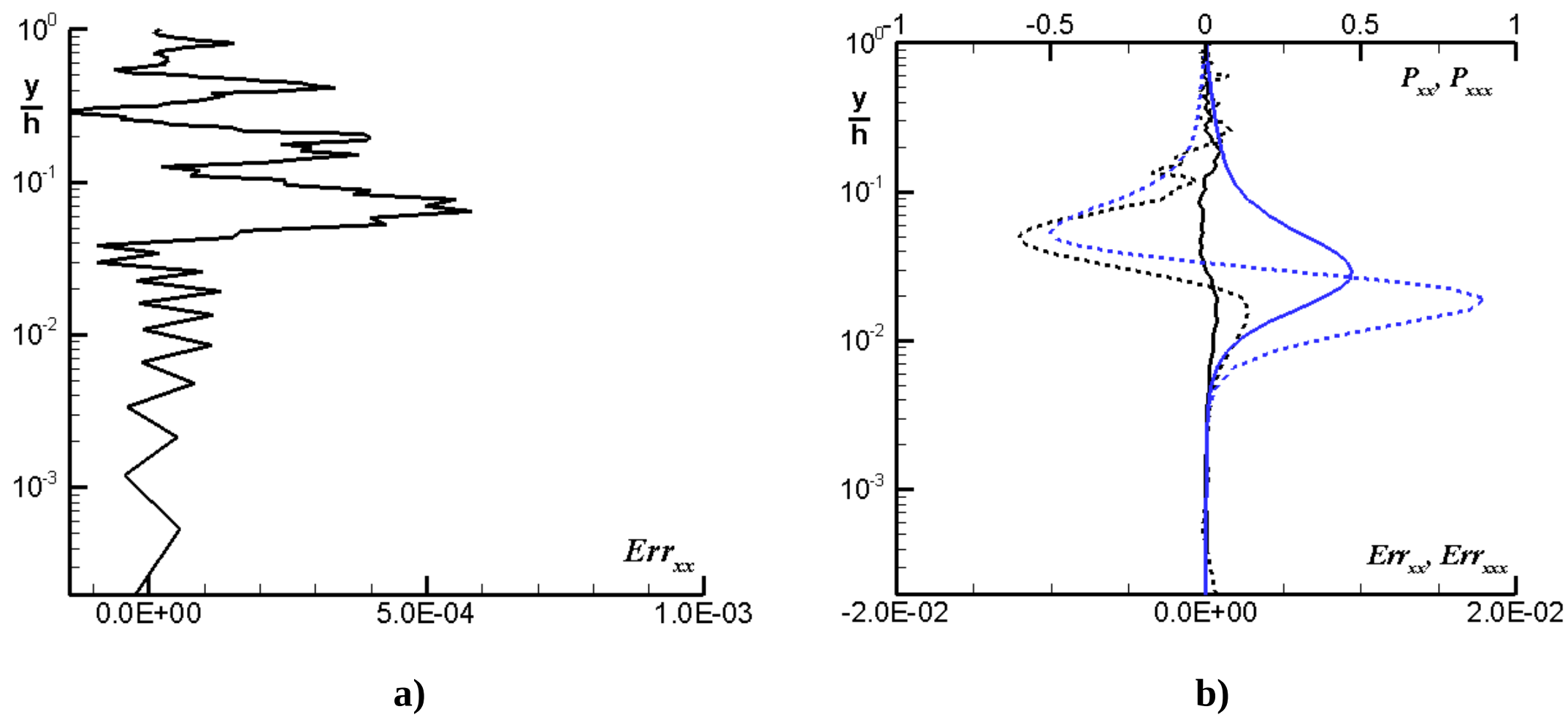

FIG. 10. The balance errors in the DNS budgets of velocity moments $\langle u^n \rangle$ in the unstrained channel flow: a) in the equation for $\langle u^2 \rangle$ at $t = 1.22 \cdot 10^5$ from [36], b) in the equations for $\langle u^2 \rangle$ and $\langle u^3 \rangle$ at $t = 5.06 \cdot 10^4$ [21]. The production terms are also shown in b) by blue lines. Solid lines correspond to the terms in the $\langle u^2 \rangle$ transport equation and dashed lines to the terms in the $\langle u^3 \rangle$ transport equation.

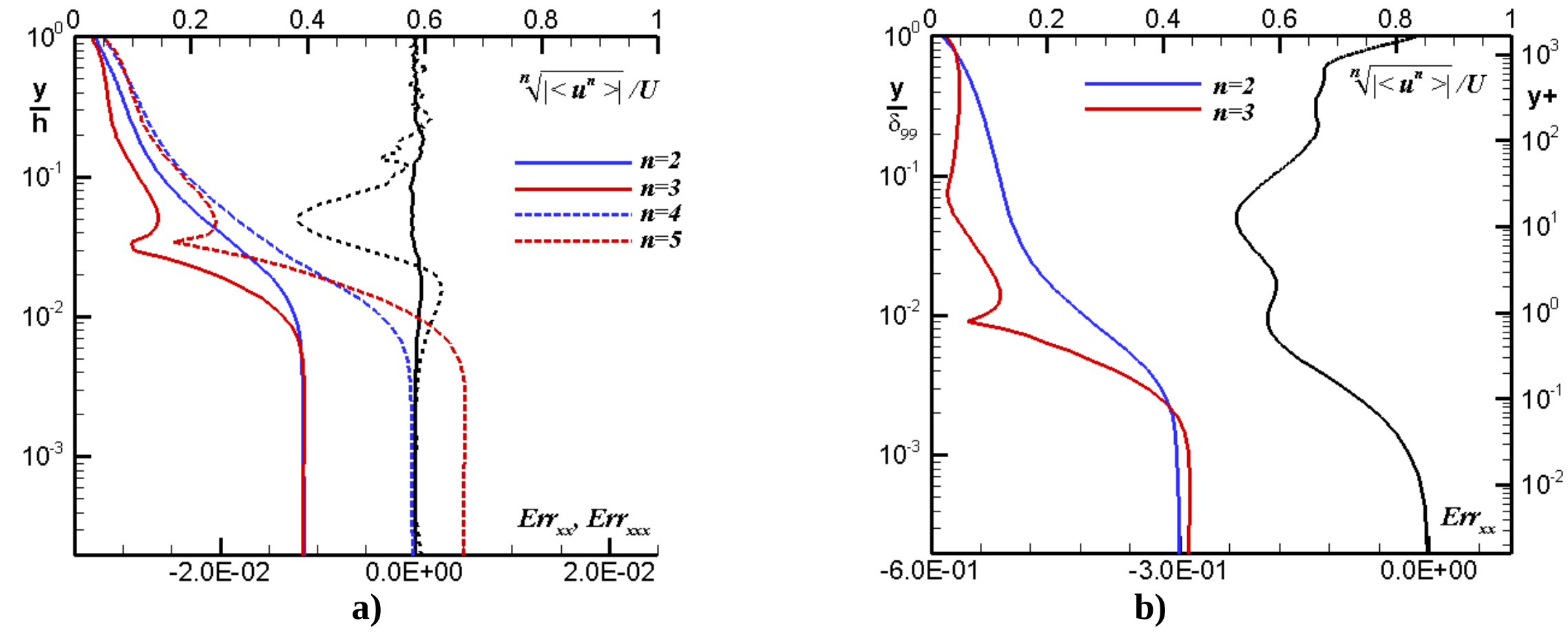

FIG. 11. Ratios between the local mean velocity and local velocity scales obtained from velocity moments $\langle u^n \rangle$, where $n = 2, \ldots, 5$ using DNS data in a) the unstrained channel [21,23], b) the zero-pressure-gradient boundary layer at $Re_\theta = 5200$ [20]. Balance errors in the budgets of $\langle u^2 \rangle$ and $\langle u^3 \rangle$ are also shown by black solid and dashed lines, respectively.

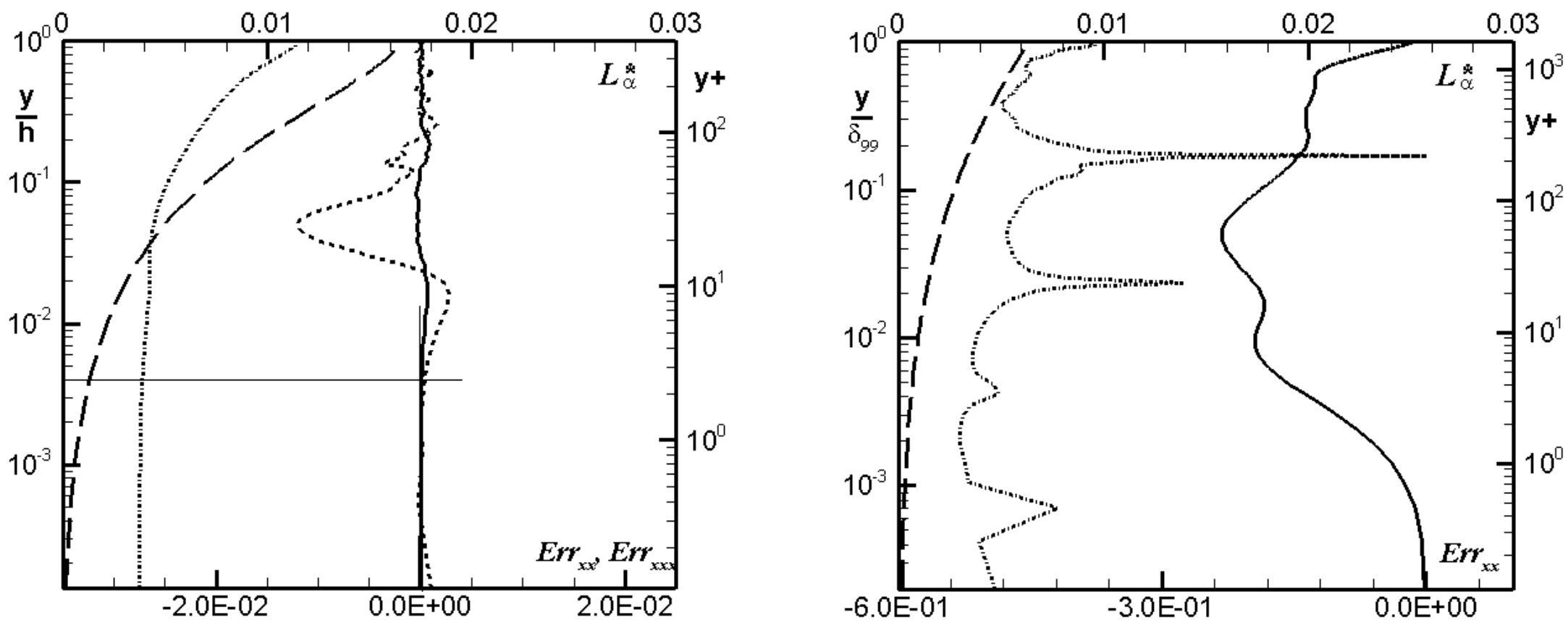

**a)** **b)**

FIG. 12. Variation of the non-dimensional lengthscales with *y*: a) $L_k^*$ (dash-dotted line) and $L_{14}^*$ (long-dashed line) obtained with DNS data for the unstrained channel from [21,23], b) a) $L_k^*$ (dash-dotted line) and $L_{13}^*$ (long-dashed line) obtained with DNS data for the zero-pressure-gradient boundary layer at $Re_\theta = 5200$ [20]. Balance errors in the budgets of $\langle u^2 \rangle$ and $\langle u^3 \rangle$ are also shown by solid and dashed lines, respectively.

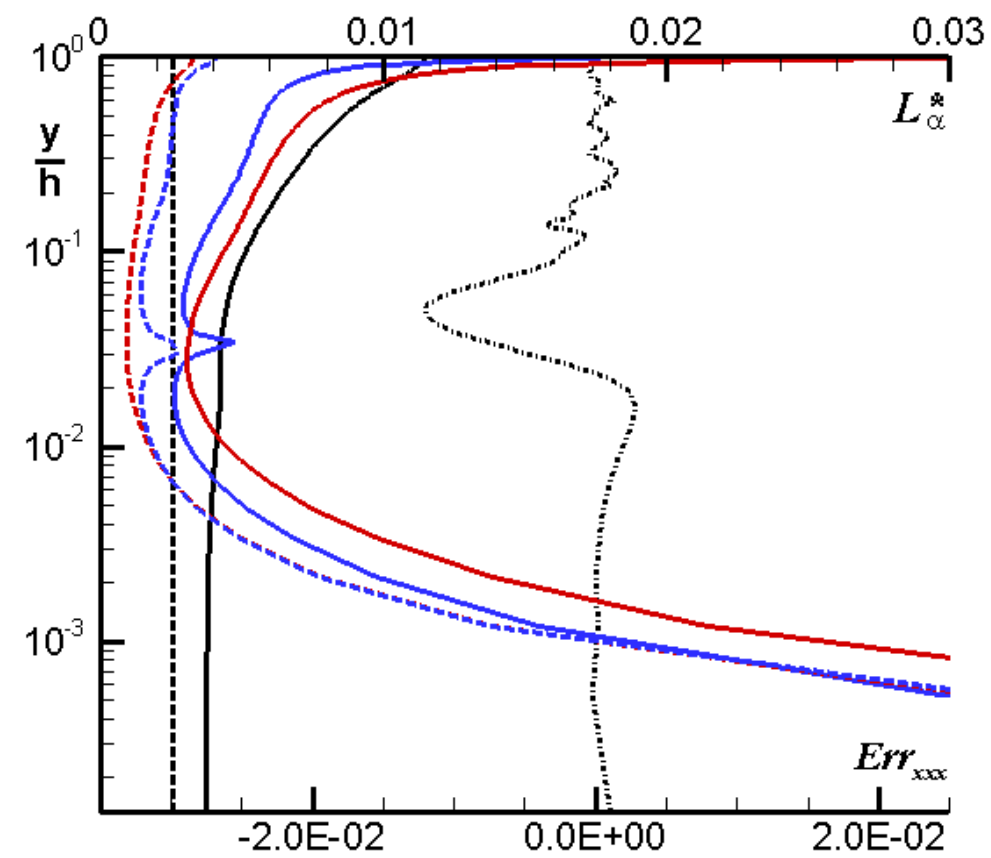

FIG. 13. Variation of the non-dimensional parameters with $y$. Lengthscales: $L_k^*$ (black solid line), ${L^*}_{inf}$ (black dashed line), ${L^*}_{p2}$ (red solid line), ${L^*}_{p3}$ (blue solid line), ${L^*}_{u2}$ (red dashed line), and ${L^*}_{u3}$ (blue dashed line). Balance errors in the budget of $\langle u^3 \rangle$ are also shown by dash-dotted line.